\documentclass[%
 reprint,
 amsmath,amssymb,
 aps,
floatfix,
]{revtex4-1}

\usepackage{graphicx}
\usepackage{dcolumn}
\usepackage{bm}
\usepackage{subcaption}
\usepackage{color}
\usepackage{float}

\begin{document}
\newcommand{\rednote}[1]{{\color{red} (#1)}}


\title{Optimizing signal recycling for detecting a stochastic gravitational-wave background
}

\author{Duo Tao}%
\email{duo.tao@ligo.org}
\affiliation{%
 Physics and Astronomy, Carleton College, Northfield, MN 55057, USA
}%

\author{Nelson Christensen}
\email{nelson.christensen@oca.eu}
\affiliation{
 Physics and Astronomy, Carleton College, Northfield, MN 55057, USA
}%
\affiliation{
 ARTEMIS, Universit\'e C\^ote d'Azur, Observatoire C\^ote d'Azur, CNRS, CS 34229, F-06304 Nice Cedex 4, France
}%

\date{\today}

\begin{abstract}
Signal recycling is applied in laser interferometers such as the Advanced Laser Interferometer Gravitational-Wave Observatory (aLIGO) to increase their sensitivity to gravitational waves. In this study, signal recycling configurations for detecting a stochastic gravitational wave background are optimized based on aLIGO parameters. Optimal transmission of the signal recycling mirror (SRM) and detuning phase of the signal recycling cavity under a fixed laser power and low-frequency cutoff are calculated. Based on the optimal configurations, the compatibility with a binary neutron star (BNS) search is discussed. Then, different laser powers and low-frequency cutoffs are considered. Two models for the dimensionless energy density of gravitational waves $\Omega_{gw}(f) = \Omega_{\alpha} (f/f_{ref})^{\alpha}$, the flat model $\alpha = 0$ and the $\alpha = 2/3$ model, are studied. For a stochastic background search, it is found that an interferometer using signal recycling has a better sensitivity than an interferometer not using it. The optimal stochastic search configurations are typically found when both the SRM transmission and the signal recycling detuning phase are low. In this region, the BNS range mostly lies between 160 and 180 Mpc. When a lower laser power is used 
the optimal signal recycling detuning phase increases, the optimal SRM transmission increases and the optimal sensitivity improves. A reduced low-frequency cutoff gives a better sensitivity limit. For both models of $\Omega_{gw}(f)$, a typical optimal sensitivity limit on the order of $10^{-10}$ is achieved at a reference frequency of $f_{ref} = 25$ Hz.
\end{abstract}

\maketitle


\section{Introduction}
The search for a stochastic gravitational wave background (SGWB) has been important in our understanding of the universe, even if the SGWB has yet to be detected~\cite{main_stochastic,PhysRevLett.118.121101,doi:10.1093/mnras/stv1538,refId0,PhysRevX.6.011035}. The Advanced Laser Interferometer Gravitational-Wave Observatory (aLIGO), one of the most sensitive scientific instruments in the world, had the capability to detect gravitational waves in its first and second observation runs~\cite{ligo_o1, gw151226, lvt151012, gw170104, gw170608, gw170814, gw170817}. Advanced Virgo~\cite{0264-9381-32-2-024001}, observing in the second observing run with aLIGO also participated in two detections~\cite{gw170814, gw170817}. In the coming observation runs, it is expected that more about the SGWB will be understood as the detectors' sensitivities improve and the LIGO-Virgo network can make deeper searches~\cite{power_law,Abbott:2017xzg}.

The introduction of the signal recycling system in aLIGO improves the sensitivity of the interferometers~\cite{aligo}. In this paper, we will show that an interferometer system with signal recycling is more sensitive for detecting the SGWB than a system without. Also, we derive the optimal configurations of the signal recycling mirror (SRM) that could achieve an optimal sensitivity for detecting a SGWB with a dimensionless energy density $\Omega_{gw}(f)$ on the order of $10^{-10}$.

It is important to note that the purpose of this paper is not to promote the use (for example by LIGO and Virgo) of signal recycling optimized for a search for a SGWB. The purpose is to understand the different optimization regimes, especially with respect to searches for compact binary systems. As will be described below, there are important and interesting differences between interferometer configurations optimized for a SGWB search and a search for gravitational waves from binary neutron star inspiral. We also find configurations where both of these searches can simultaneously effectively occur.

In this study the sensitivities of the Advanced LIGO interferometers are simulated with the the Gravitational Wave Interferometer Noise Calculator (GWINC)~\cite{gwinc}; the sensitivities are assumed to be the same for the two detectors. This is the software that has been used to produce predictions of the performance of Advanced LIGO~\cite{1742-6596-610-1-012013,aligo,Evans-Mathworks}. The signal recycling results presented here were generated with GWINC based on the analysis of Buonanno and Chen~\cite{quantum_noise,PhysRevD.69.102004}. 

Section II presents the general approach to optimize the SGWB search and the result for a particular power usage and frequency cutoff. Section III considers the optimization under different laser powers and low-frequency cutoffs. Section IV considers the $\Omega_{gw}(f) \propto f^{2/3}$ power-law model~\cite{power_law} for the SGWB. Section V summarizes the results of this project and considers possible topics of future work.

\section{Optimization for the SGWB Search}
\subsection{Signal Recycling Parameters}
\begin{figure*}[t]
  \centering
    \includegraphics[width=0.7\textwidth]{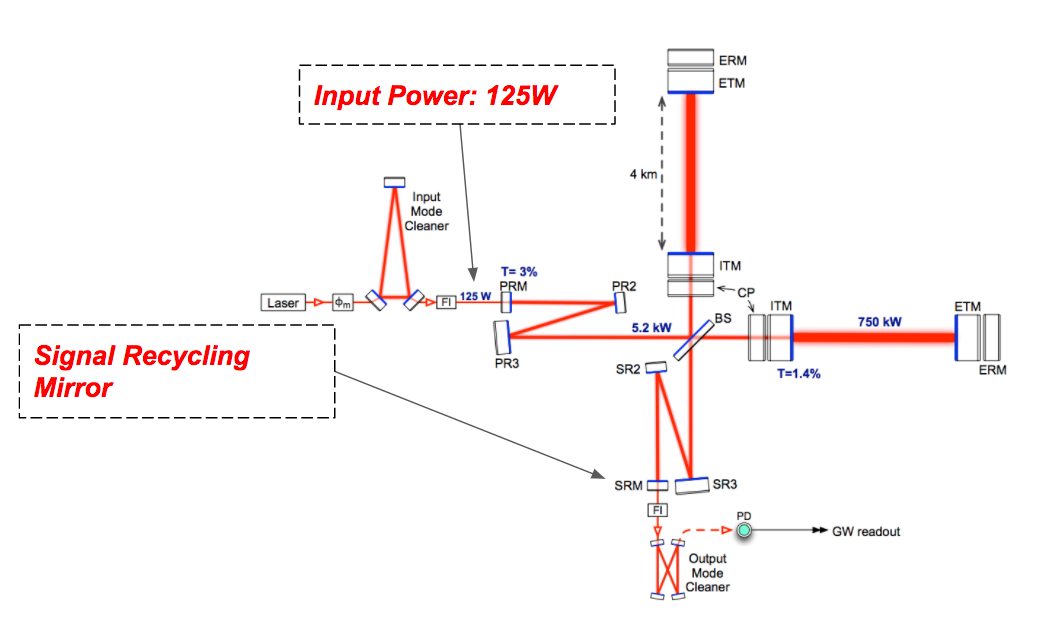}
    \caption{The aLIGO optical configuration~\cite{aligo}. The SRM is shown and the input power at the power recycling mirror is assumed for the first part of this analysis to be 125 W.}
    \label{fig:srm}
\end{figure*}

The aLIGO optical configuration is displayed in Fig.~\ref{fig:srm}~\cite{aligo}. The SRM is a mirror at the output port of the interferometer, sending part of the signal (output beam) back into the interferometer to create resonances at certain frequencies in order to amplify the signal sensitivity~\cite{PhysRevD.38.2317,thesis}.

We are interested in two parameters pertaining to the signal recycling configuration, the transmission of the SRM and the signal recycling detuning phase. It should be noted that the transmission is the light power transmission, the percentage of power transmitted through the mirror, in contrast to the amplitude transmission. It is given as~\cite{quantum_noise}
\begin{equation}\label{eqn:transmission}
T = 1 - R - \lambda_{SR}
\end{equation}
where $R$ is the reflectivity, $T$ is the transmission and $\lambda_{SR}$ is the loss inside the signal recycling cavity (SRC). This is the sum of mismatch loss and beamsplitter loss. In this work, we assume that the mismatch loss is zero. This is the mismatch between the modes in the arms and the SRC.
The beamsplitter loss is assumed to be $2\times10^{-3}$. 
It should be stressed that the improvement derived from the use of signal recycling is very sensitive to these losses~\cite{PhysRevD.69.102004}. In this present study we strive to mimic previous calculations predicting the performance of Advanced LIGO~\cite{gwinc,1742-6596-610-1-012013,aligo}.
Consequently, $T$ could not be greater than 99.8\% unless $T=1$ when there is no signal recycling.
Another parameter considered in this work is the signal recycling detuning phase~\cite{gwinc,quantum_noise}. Note that Advanced LIGO, so far in its initial observing runs, has operated in a broadband mode where the signal recycling detuning phase has been set to zero~\cite{aligo}.

\subsection{Sensitivity limit $\Omega_{gw}(f)$}
A gravitational wave makes a slight modification to flat space. Using linearized general relativity 
\begin{equation}
g_{\mu \nu} \approx \eta_{\mu \nu} + h_{\mu \nu} ~ ,
\end{equation}
where $g_{\mu \nu}$ is the spacetime metric, $\eta_{\mu \nu}$ is the flat space Minkowski metric, and $h_{\mu \nu}$ is the perturbation to the metric; in this case $h_{\mu \nu}$ is the gravitational wave. Gravitational waves have two possible polarizations, and like electromagnetic waves, their effect is transverse to the direction of propagation.

Gravitational wave detectors, like Advanced LIGO and Advanced Virgo, were constructed to measure a gravitational wave strain, with amplitude $h(t)$ (a dimensionless quantity). In reality, a gravitational wave detector will produce a signal, $s(t)$, that is the sum of the gravitational wave stain, $h(t)$, and noise, $n(t)$, namely
\begin{equation}
s(t) = h(t) + n(t) ~,
\end{equation}
where $n(t) >> h(t)$. The SGWB will just appear as noise in a single detector. The way to detect the SGWB is therefore to use two detectors, and perform a correlation on the two output signals. Imagine two co-located detectors, subjected to the same gravitational wave $h(t)$, then
\begin{equation}
<s_{1}(t) s_{2}(t)> ~ = ~ <(n_{1}(t) + h(t))(n_{2}(t) + h(t))> ~ ,
\end{equation}
where the brackets indicate a time average. If the noise in each detector is statistically independent from one another, and also with the SGWB, then
\begin{equation}
<s_{1}(t) s_{2}(t)> ~ \approx ~ <h(t) h(t)> ~ .
\end{equation}
In order for there to be no correlated noise, the detectors should be displaced from one another~\cite{PhysRevD.91.022003}. The two Advanced LIGO detectors are separated by 3000 km. However, even with substantial separation it has been demonstrated that global magnetic field noise from the Schumann resonances can have a coherence between the LIGO and Virgo sites~\cite{PhysRevD.87.123009,0264-9381-34-7-074002}. For this present study we will ignore these magnetic field correlations. Because of the distance separation and the fact that the two interferometers are not perfectly aligned with respect to one another, extracting the common SGWB signal from the correlation between the two detectors becomes more complicated; see~\cite{nelson1993,main_stochastic,Romano:2016dpx} for explicit details on how this is accomplished for LIGO and Virgo.

The correlation between the outputs of two gravitational wave detectors, located at $\vec{x_1}$ and $\vec{x_2}$, will be proportional to the root mean square (rms) of the gravitational wave strain, $h^{2}_{rms}$. This can be related to the one-sided spectral density of the gravitational wave, $S_{h}(f)$, or
\begin{equation}
h^{2}_{rms} = \Bigg \langle \sum_{i,j} h_{ij} h_{ij} \bigg \rangle = \int_{0}^{\infty} df S_{h}(f) ~ .
\end{equation}
$S_{h}(f)$ can then be related to the gravitational wave energy density,
\begin{equation}
\rho_{gw} = \int_{0}^{\infty} df S_{h}(f) \frac{\pi c^{2} f^{2}}{8 G} ~ .
\end{equation}
We can then relate the energy density of the SGWB, $\rho_{gw}$, and the energy density  per unit frequency, $\rho_{gw}(f)$,
\begin{equation}
\rho_{gw} = \int \rho_{gw}(f) df 
\end{equation}
Noting that $\rho_{gw}(f)$ is a function of the frequency of the gravitational waves, we can then define another quantity $\Omega_{gw}(f)$ in Eq.~\ref{eqn:omega_def} as~\cite{nelson1993}
\begin{equation}\label{eqn:omega_def}
\Omega_{gw}(f) = \frac{1}{\rho_c} \frac{d \rho_{gw}}{d\ln{f}} = \frac{f\rho_{gw}(f)}{\rho_c}.
\end{equation}
$\Omega_{gw}(f)$ is the energy density of the SGWB per logarithmic frequency interval, normalized by $\rho_c$, the closure density of the universe. We can express a frequency dependence of the energy density by writing 
\begin{equation}
\Omega_{gw}(f) = \Omega_{\alpha} (f/f_{ref})^{\alpha}
\label{eq:pl_general}
\end{equation}
where $f_{ref}$ is a reference frequency. Based on this, it is found that the detection sensitivity limit is given as ~\cite{thesis, nelson1993}
\begin{equation}\label{eqn:lim_pl}
\Omega_\alpha\geq\frac{25\pi c^2}{16\rho_c G}\sqrt{\frac{2}{T}}\Big[\int \frac{\gamma^2(\vec{x_1},\vec{x_2},f)}{h_n^4(f)f^{6 - 2\alpha}f_{ref}^{2\alpha}}df\Big]^{-1/2},
\end{equation}
where $\gamma(\vec{x_1},\vec{x_2},f)$ is the overlap reduction function~\cite{nelson1993}; this quantity takes into account the distance separation between two detectors, and their relative orientation~\cite{Romano:2016dpx}.  $h_n(f)$ is the interferometers' noise spectrum (assumed to be the same for the two detectors used in the correlation experiment) and $T$ is the integration time, which in this case is assumed to be a year. 
For most of the work in this study we assume $\Omega_{gw}(f)$ is a constant for SGWB, i.e. $\alpha = 0$, and thus we have
\begin{equation}\label{eqn:bg_flat}
\Omega_{gw}(f) = \Omega_0.
\end{equation}
When $\Omega_{gw}(f)$ is a constant, according to Eq.~\ref{eqn:lim_pl}, the detection limit $\Omega_0$ is 
\begin{equation}\label{eqn:bb_omega}
\Omega_0\geq\frac{25\pi c^2}{16\rho_c G}\sqrt{\frac{2}{T}}\Big[\int \frac{\gamma^2(\vec{x_1},\vec{x_2},f)}{h_n^4(f)f^6}df\Big]^{-1/2}.
\end{equation}
While a constant energy density would diverge for an integral over all frequencies, for the present study we are only concerned with the frequency dependence of the SGWB energy density over the bandwidth of the detectors.

According to Eq.~\ref{eqn:bb_omega}, the sensitivity to a SGWB will be best at low frequencies for aLIGO because the overlap reduction function is larger there, plus the $f^{-6}$ dependence; certainly the frequency dependence of the interferometers' noise $h_n(f)$ will also play an important part in the sensitivity to a SGWB.
The computation is done by GWINC~\cite{gwinc}.
\subsection{The method}
\begin{table*}[t]
  \centering
  \caption{Parameters for the four configurations involved in the comparison. The BNS ranges and the $\Omega_{0}$ are calculated assuming an input laser power of 125 W and low-frequency cutoff 10 Hz.}
  \begin{tabular}{c|c|c|c|c}
    Configuration & SRM transmission & signal recycling detuning phase (degrees) & $\Omega_{0}$ & BNS Range(Mpc) \\
    \hline
    No recycling & 1 & 0 & $1.1 \times 10^{-8}$ & 116.6 \\
    Optimal BNS & 0.2 & 16 & $2.3 \times 10^{-9}$ & 207.9 \\
    Optimal SGWB & 0.015 & 2.7 & $6.8 \times 10^{-10}$ & 154.4 \\
    Nominal & 0.33 & 0 & $2.3 \times 10^{-9}$ & 191.2
  \end{tabular}
  \label{tab:conf}
\end{table*}
We calculate all the sensitivity results for configurations with signal recycling detuning phase from $-180^\circ$ to $180^\circ$ and SRM transmission from 0\% to 100\%. Note that we would abandon any results greater than 99.8\% other than 100\% so as to avoid conflict with Eq.~\ref{eqn:transmission} but it turns out that none of the optimization results produced in this work has transmission between 99.8\% and 100\%. Therefore, we search for the optimal sensitivity in a 2D domain of $[transmission \times phase]$. The step size through the parameter space determines the accuracy of the optimization. Using smaller steps would improve the accuracy of the optimization results. However, computational resources limit the size of the steps such that if the steps are too small, there will be too many candidate configurations produced and the amount of computations would be unfeasible. Therefore, we take the following two-step approach to obtain a finer optimization without too much computational burden.
\begin{enumerate}
\item We start with a full scan over the whole region with transmission resolution of $0.1\%$ and phase resolution of $1^\circ$. Transmission ranges from $0\%$ to $100\%$ and the phase ranges from $-180^\circ$ to $180^\circ$. Therefore, there will be $1000\times 180=180000$ configurations considered and the accuracy of the results is $0.1\%$ and $1^\circ$.
\item Then, we do a localized scan within the accuracy limit around the optimized configurations we get in the first step. Suppose that in the first step we find the transmission $T$ and the phase $\phi$ that achieve the best sensitivity. We then scan the 2D domain $[T-0.1\%, T+0.1\%]\times [\phi - 1^\circ, \phi + 1^\circ]$ with a transmission resolution of $0.01\%$ and a phase resolution of $0.1^\circ$, both ten times finer than the first step. Thus, the second step involves $20\times 20=400$ configurations in total.
\end{enumerate}
Hence, the method above gives results with accuracy of $0.01\%$ in transmission and $0.1^\circ$ in phase.

We are aware that this method has its shortcomings such that in the second step we might be approximating a local extremum. If there are multiple local extrema and in the first step we locate a local extremum that is not the global extremum due to insufficient resolution, the second step will only get us closer to the local extremum. The problem of insufficient resolution always exists when we optimize a function numerically. However, we shall explain at the end of this section that this might not be a significant problem.

We have examined using alternative optimization methods such as gradient descent or a simplex search to find the optimal transmission and phase. The benefit of using a more advanced optimization algorithm is the time efficiency. One would use them when it is beyond the computing capability to enumerate all possible inputs. Fortunately, in this present study, both the transmission and the phase have a limited range. We can find optimal values with satisfactory accuracy using a grid search. Therefore, compared to the alternatives, the grid search gives us the full picture of the behavior, finds the global optimum to a satisfactory accuracy without overwhelming our computational ability, and does not have much technical risk due to its simplicity. Therefore, the grid search is our most feasible choice in this particular situation.

\subsection{Optimization results}
We assume 125 W incident on the power recycling mirror.
Using the method described above, it is found that $1.5\%$ transmission and $2.7^\circ$ of signal recycling detuning phase give us the optimal SGWB sensitivity with the limit of $\Omega_{0}$ being $6.8 \times 10^{-10}$. Due to the fact that the smallest grid size used in the second step to locate this optimum configuration is $0.01\%$ for transmission and $0.1^\circ$ for the signal recycling detuning phase, the uncertainties in the results are $0.01\%$ for transmission and $0.1^\circ$ for the signal recycling detuning phase. Taking the uncertainties into account, the results for the optimization of an interferometer with 125 W of laser power input for the SGWB search are $1.5\%\pm 0.01\%$ for transmission, and $2.7^\circ\pm0.1^\circ $ for the signal recycling detuning phase.

\begin{figure}[t]
	\includegraphics[width=0.45\textwidth]{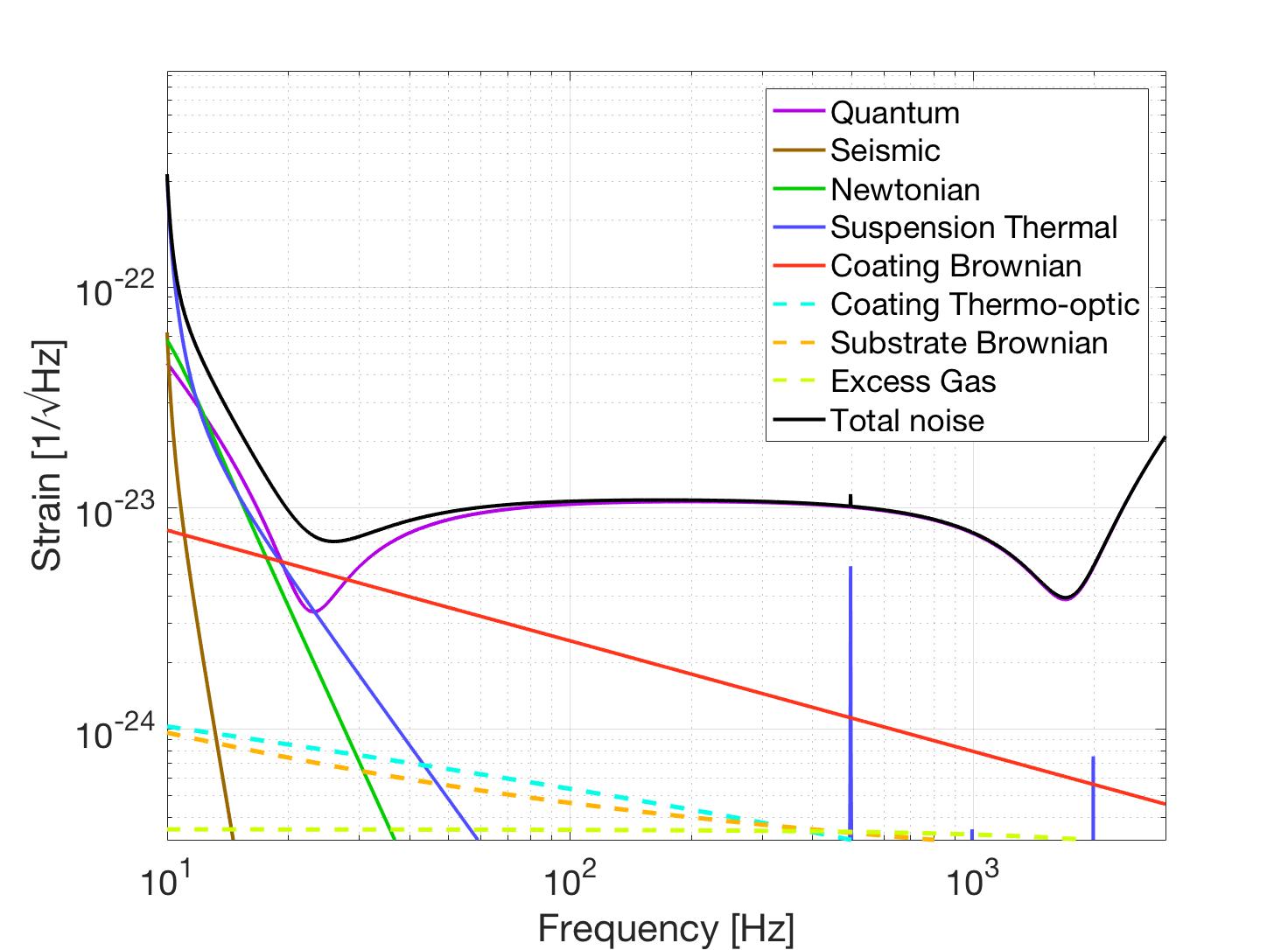}
	\caption{The noise spectrum for the optimal SGWB configuration at 125W, with the low-frequency cutoff as 10Hz. The signal recycling detuning phase is $2.7^{\circ}$ and the transmission is $1.5\%$~\cite{gwinc}. }
	\label{fig:opt_spectrum}
\end{figure}

The optimal noise spectrum for 125 W is shown in Fig.~\ref{fig:opt_spectrum}. We can see a dip in the low frequency range around 25 Hz, which should explain why this configuration does well for SGWB search. The $f^{-6}$ term in Eq.~\ref{eqn:bb_omega} gives the low frequency noise spectrum much more weight than the high frequency part so a low frequency improvement is more effective to enhance the SGWB sensitivity. 

\subsection{Comparison with other configurations}
\begin{figure}[t]
    \centering
    \includegraphics[width=0.45\textwidth]{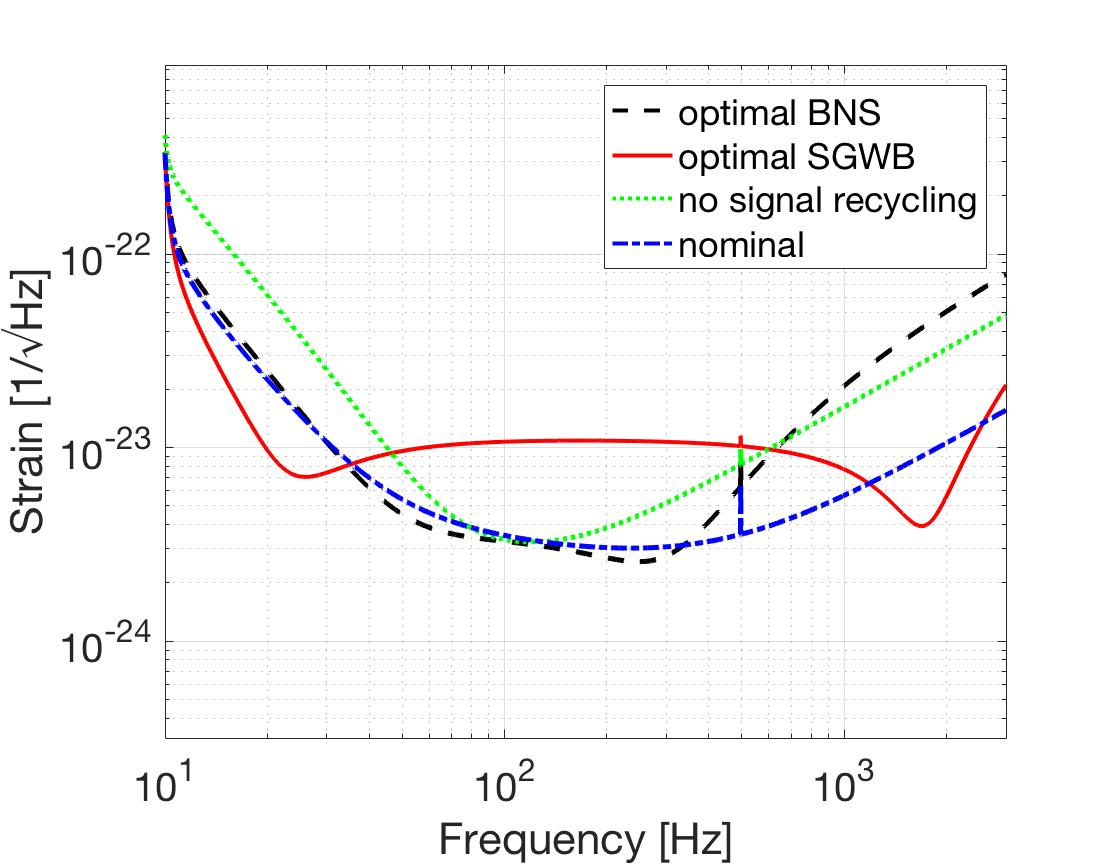}
    \caption{Comparison of the detector sensitivity optimized for a SGWB search with other configurations: no signal recycling (transmission = 1, phase = 0), BNS optimized (transmission = 20\%, phase = $16^{\circ}$) and nominal configuration (transmission = 33\% and phase = 0).}
    \label{fig:spec_comp}
\end{figure}

The total noise spectrum of a SGWB optimized configuration is compared with other configurations in Fig.~\ref{fig:spec_comp}. There are three other configurations with which we have interest: no signal recycling, the configuration optimized for a binary neutron star (BNS) search and the nominal configuration.
\begin{enumerate}
\item \textbf{No signal recycling}: when there is no signal recycling installed in the interferometer, all light passes so we set the transmission to one. The noise spectrum is shown in Fig.~\ref{fig:spec_comp}. The sensitivity limit for $\Omega_{0}$ is $1.1 \times 10^{-8}$.

\item \textbf{Optimal BNS search}: this configuration (transmission $20\%$ and detuning phase $16^\circ$) has an optimal BNS range of 207.9 Mpc ~\cite{aligo}. The noise spectrum is shown in Fig.~\ref{fig:spec_comp}. We can see that its optimization is different than the optimal SGWB spectra in Fig.~\ref{fig:opt_spectrum}. The former has its best sensitivity around mid-frequency 100 Hz while the latter benefits the most from low-frequency improvements. The sensitivity limit for $\Omega_{0}$ is $2.3 \times 10^{-9}$.

\item \textbf{Nominal configuration:} the nominal configuration (transmission $33\%$ and phase zero) has a flat and smooth noise spectrum, displayed in Fig.~\ref{fig:spec_comp}. The sensitivity limit for $\Omega_{0}$ is $2.3 \times 10^{-9}$.
\end{enumerate}

From all of the three comparison plots presented in Fig.~\ref{fig:spec_comp}, it can be seen that the optimal SGWB configuration has a significant improvement at low frequencies. At the same time, there is a sacrifice of mid-frequency performance around 100 Hz. The low-frequency sensitivity is more important for a SGWB search, hence the better sensitivity for $\Omega_{0}$. 

Parameters for the optimal SGWB search and the other three configurations are summarized on Table~\ref{tab:conf}. The SGWB optimized configuration is about three times more sensitive than the BNS optimized configuration for SGWB search. Also, the BNS optimized configuration has a BNS range 53 Mpc further than the BNS range of the SGWB configuration. If we were to choose one configuration to use for both the SGWB search and the BNS search among the four configurations in Table~\ref{tab:conf}, the BNS optimized configuration would probably be the preferred choice since it has the best BNS range and a reasonable SGWB $\Omega_{0}$ limit.

\subsection{Compatibility with BNS search}
In this section, we will address the question as to whether there is a particular signal recycling configuration such that a SGWB search could be compatible with a BNS search. There are two approaches used here.

The first approach is quantitative. We are looking for a configuration that has good sensitivities for both the SGWB search and the BNS search. Based on this, we can limit the scope to configurations having nominally good sensitivities for the SGWB search. Thus, we list all configurations that give $\Omega_{0} < 10^{-9}$ and calculate the maximum BNS range among them. It is found that the maximum BNS range is 165 Mpc. Compared with the 207.9 Mpc for the BNS optimized configuration and 191 Mpc for the nominal configuration, these configurations are significantly less effective for a BNS search.

The second approach is qualitative. We plot all $\Omega_{0}$ and BNS ranges on a 2D region of the $[transmission \times detuning phase]$ space. The plots are shown in Fig.~\ref{fig:scan}. Both the SGWB search and the BNS search are optimized on the center of the left edge where the phase and transmission are low. However, the darkest region of the SGWB plot is even closer to the origin than the optimal (brightest) region of the BNS range. The optimal configuration of the SGWB search has even lower transmission and phase. It also deteriorates more quickly as the transmission and phase increases. When it gets close to the high BNS range region, the sensitivity rapidly worsens.

\subsection{Issues with potential extrema}
\begin{figure*}[t]
    \centering
    \begin{subfigure}[t]{0.5\textwidth}
        \centering
        \includegraphics[width=\textwidth]{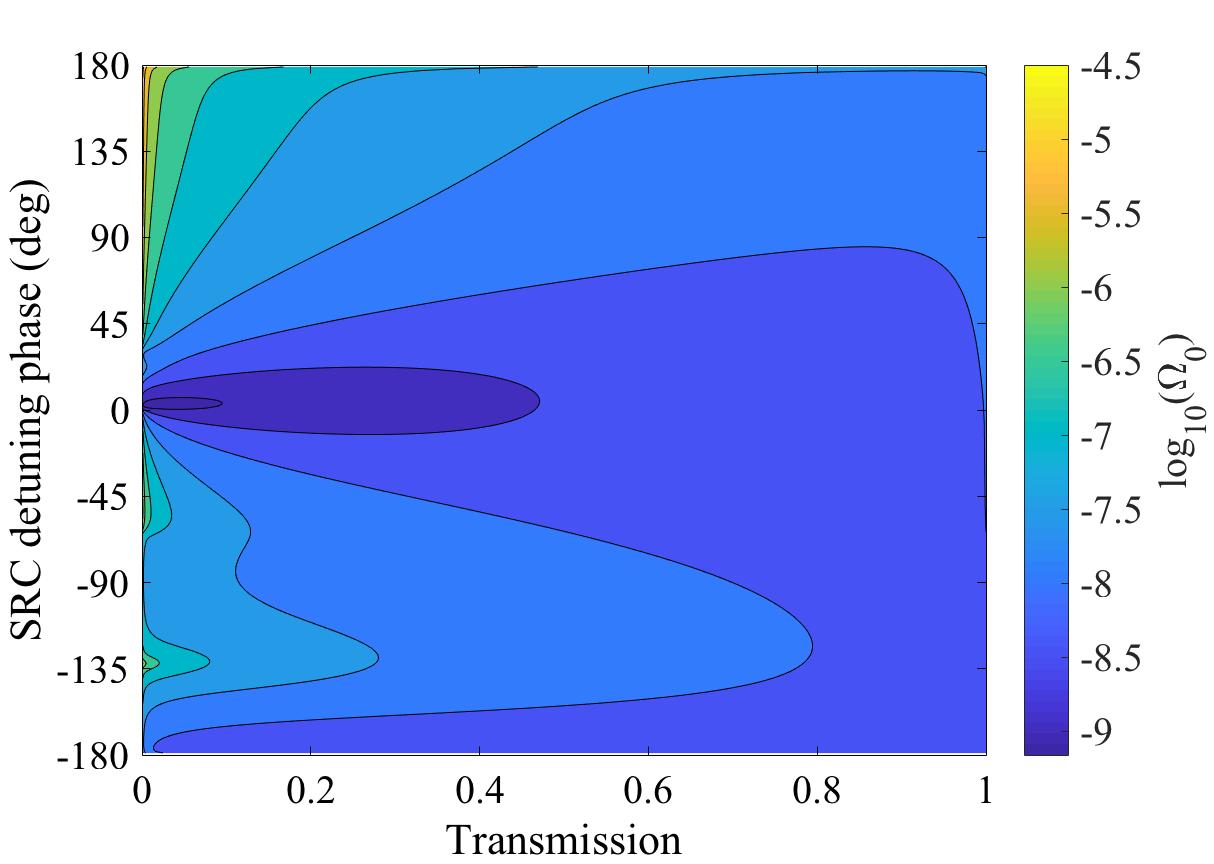}
    \end{subfigure}%
    \begin{subfigure}[t]{0.5\textwidth}
        \centering
        \includegraphics[width=\textwidth]{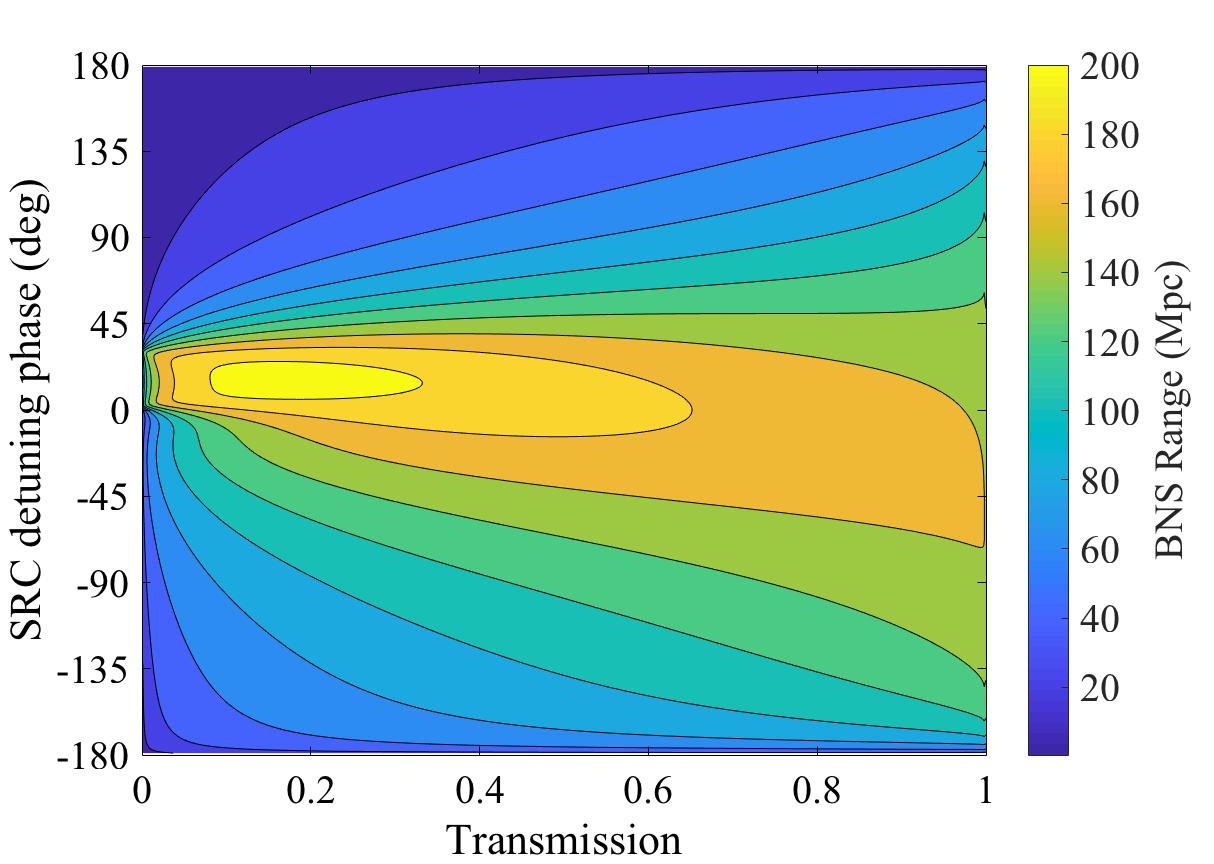}
    \end{subfigure}
    \begin{subfigure}[t]{0.5\textwidth}
        \centering
        \includegraphics[width=\textwidth]{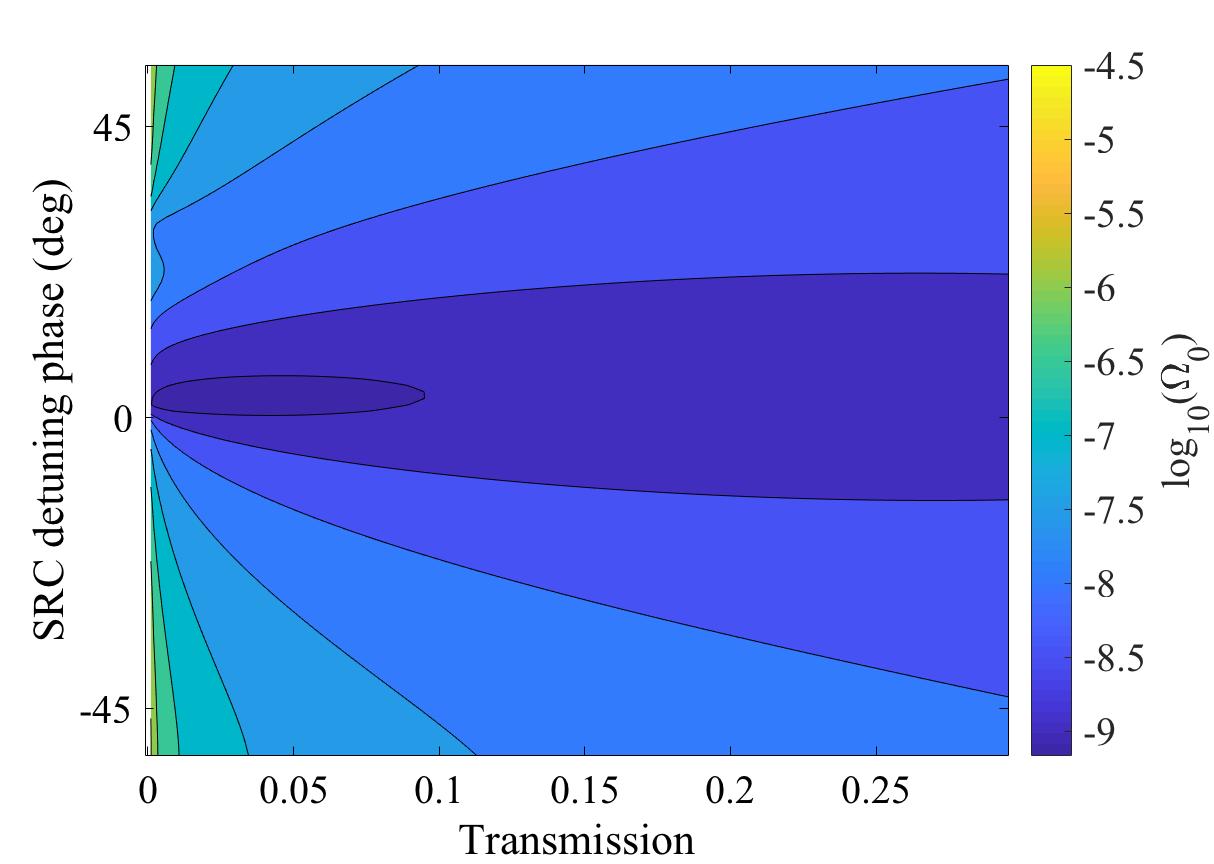}
    \end{subfigure}%
    \begin{subfigure}[t]{0.5\textwidth}
        \centering
        \includegraphics[width=\textwidth]{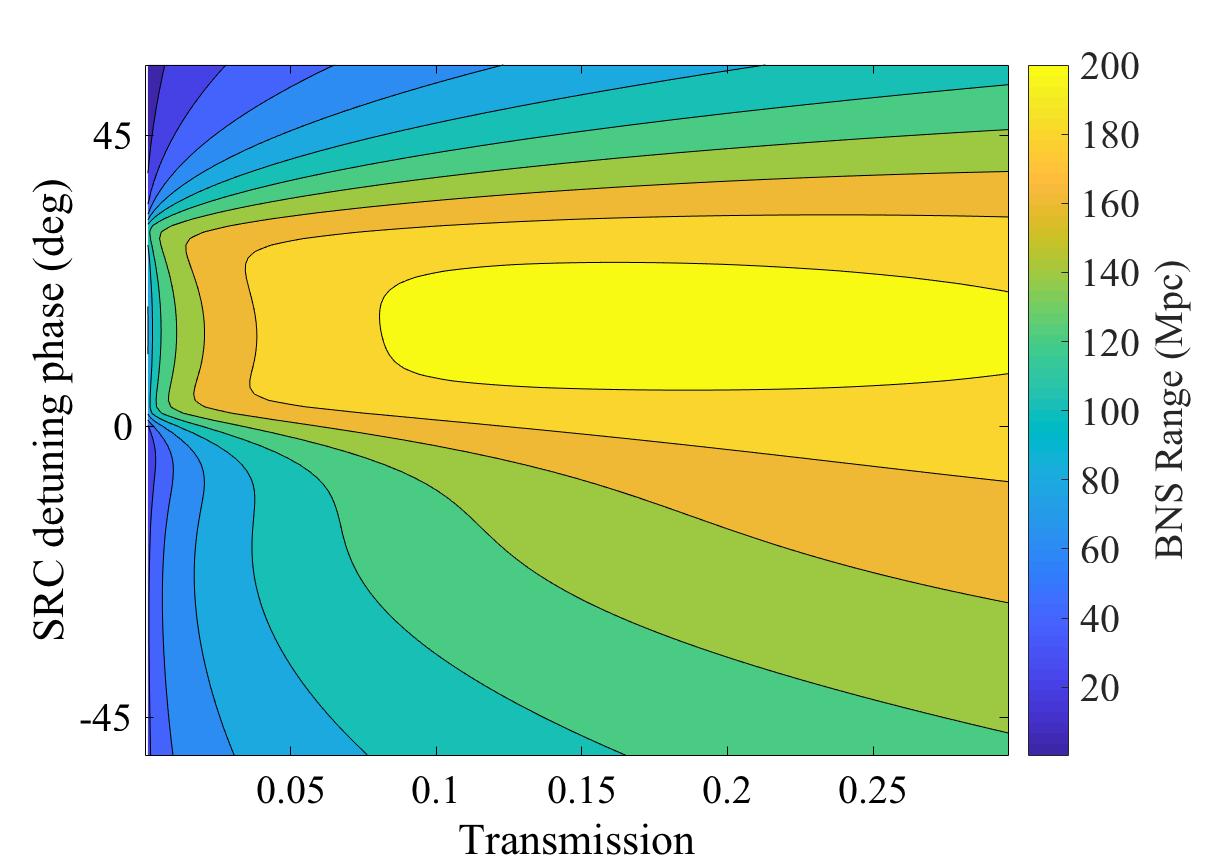}
    \end{subfigure}
    \caption{Plots of $log_{10}(\Omega_{0})$ and the BNS range as functions of SRM transmission and signal recycling detuning phase. These calculations assume a laser power input of 125 W and low-frequency cutoff of 10 Hz. On the left are plots of $log_{10}(\Omega_{0})$ and on the right are plots of BNS range. We can see that both are optimized in the low-phase low-transmission region close to the origin. However, the optimal region for $\Omega_{0}$ is smaller and closer to the origin than the optimal region of the BNS range. This is clear in the zoomed-in plots on the bottom.}
    \label{fig:scan}
\end{figure*}

With the information in Fig.~\ref{fig:scan}, we are now able to address the issue left in part C. In part C, we were worried about the validity of our method such that there might be a global optimal configuration between the smallest steps of our scanning of the phase and transmission. However, examining the colors in Fig.~\ref{fig:scan}, we notice that there is not another region that has $\Omega_{0}$ comparable to the darkest region around the origin. Thus, since we located the extremum in the low-phase and low-transmission region, it is unlikely for other local extrema to compete with it.
\section{Using a Different Laser Power and Low-frequency Cutoff}
For all the work above, we have assumed a laser power of 125 W and low-frequency cutoff of 10 Hz. In this section, we will optimize the signal recycling parameters using different laser powers and low-frequency cutoffs, while keeping all of the other aLIGO (target sensitivity) noise terms the same.

\subsection{Optimal configurations}
\begin{figure}[t]
    \centering
    \begin{subfigure}[t]{0.4\textwidth}
        \centering
        \includegraphics[width=\textwidth]{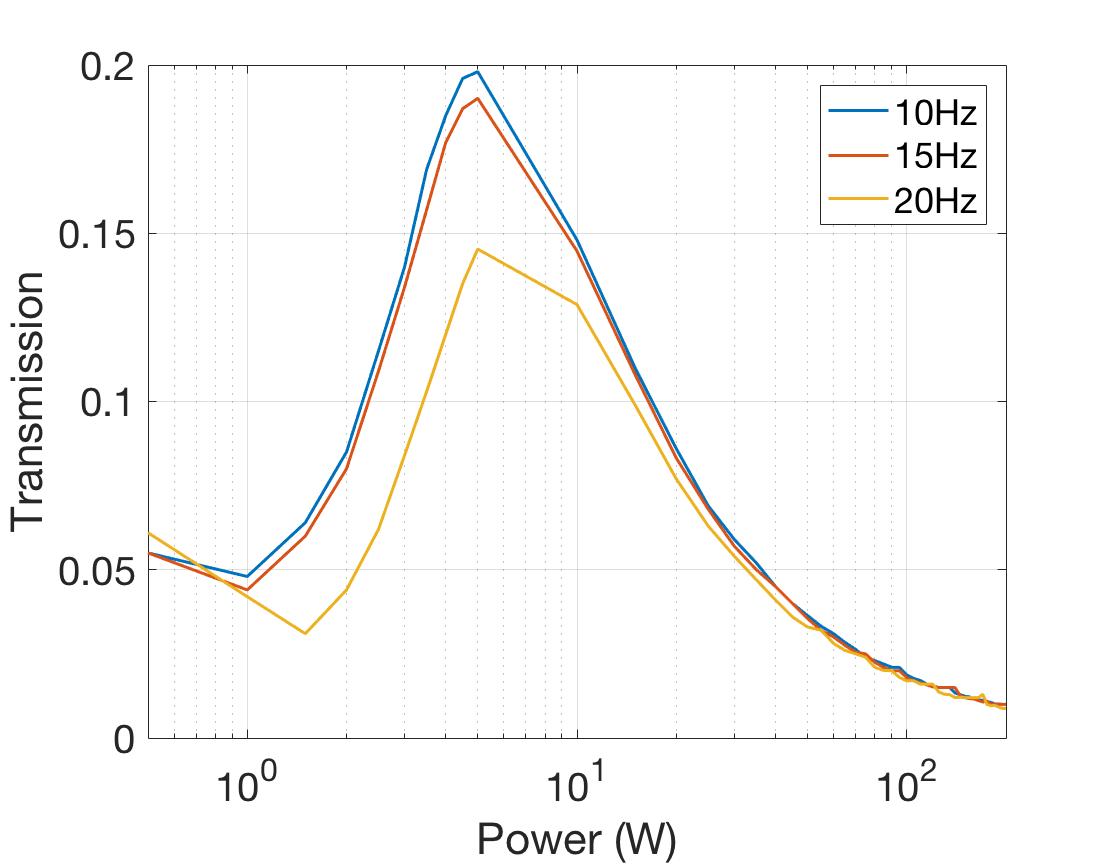}
    \end{subfigure}\\
    \begin{subfigure}[t]{0.4\textwidth}
        \centering
        \includegraphics[width=\textwidth]{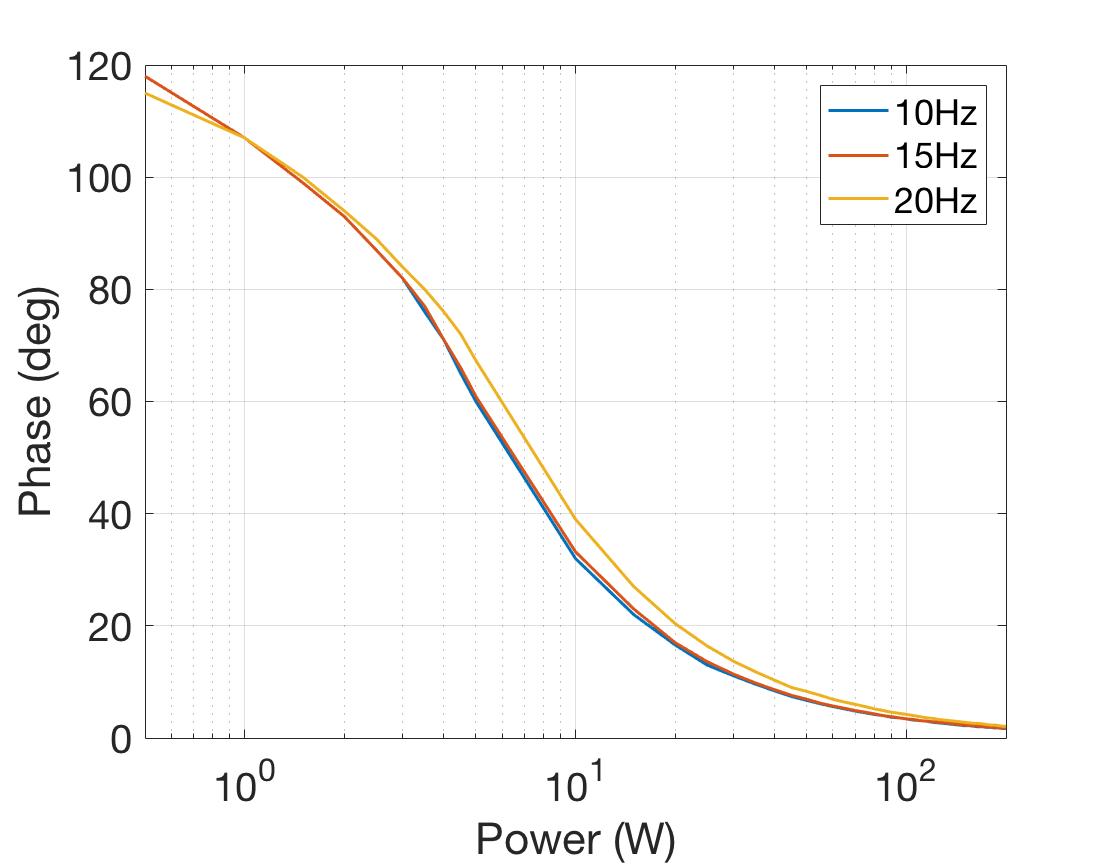}
    \end{subfigure}
    \caption{Optimal SRC configurations as a function of input laser power. The plot shows the SRM transmission and the signal recycling detuning phase that produce the lowest $\Omega_{0}$ limit, for input laser powers between 0.5 W and 200 W. The top plot shows that if we increase the input power, we should increase the transmission when the laser power is below 5 W, but decrease the transmission above 5 W. The bottom plot shows that when we increase the input laser power, we should decrease the detuning phase, in order to produce the best sensitivity to SGWB.}
    \label{fig:pow_config}
\end{figure}

The relation between power and the optimization is investigated in two groups: low power below 5 W and high power above 5 W up to 200 W. We use a smaller step size for lower laser power. For the high power group, we use a step size of 5 W. Namely, we optimize all interferometers with laser powers from 5 W to 200 W, in steps of 5 W. For the low power, we use a step size of 0.5 W. Thus, we optimize all interferometers with laser powers from 0.5 W up to 5 W in steps of 0.5 W. For each laser power, three low-frequency cutoffs, 10 Hz, 15 Hz and 20 Hz, are considered. Therefore, combining 49 input powers with three low-frequency cutoffs for each power, the signal recycling of 147 possible interferometer parameters are optimized for the SGWB search. We find the optimal signal recycling configurations (i.e. the SRM transmission and the signal recycling detuning phase that produce the lowest $\Omega_{0}$ limit) for all the 147 input power and low-frequency cutoff combinations. The results are shown in Fig.~\ref{fig:pow_config}; note that this figure shows the results for the three options for the low-frequency cutoff: 10 Hz, 15 Hz and 20 Hz.  

The top plot of Fig.~\ref{fig:pow_config} shows the SRM transmission that produces the lowest $\Omega_{0}$ limit as a function of laser power. If we increase the input laser power, we should increase the SRM transmission if the input power is below 5 W but decrease the SRM transmission if the input power is above 5 W. For example, with a 10 Hz cutoff, if we increase the laser power from 0.5 W to 5 W, we should increase the SRM transmission from 5.5\% to 19.8\% to achieve the lowest $\Omega_{0}$ limit. However, if we keep increasing the input power from 5 W to 200 W, we find the SRM transmission that produces the lowest $\Omega_{gw}$ limit decreases from $19.8\%$ to $1.0\%$. 

The bottom plot of Fig.~\ref{fig:pow_config} shows the signal recycling detuning phase of the optimal configuration as a function of the input power. We should decrease the signal recycling detuning phase if we increase the input power. For example, if the input power is 0.5 W, the detuning phase that gives the lowest $\Omega_{0}$ is 118$^\circ$. If we instead use 200 W laser, we should decrease the phase to 1.7$^\circ$. According to Fig.~\ref{fig:pow_config}, the phase goes down to zero. 



\subsection{Optimal $\Omega_{0}$}
\begin{figure}[b]
	\includegraphics[width=0.45\textwidth]{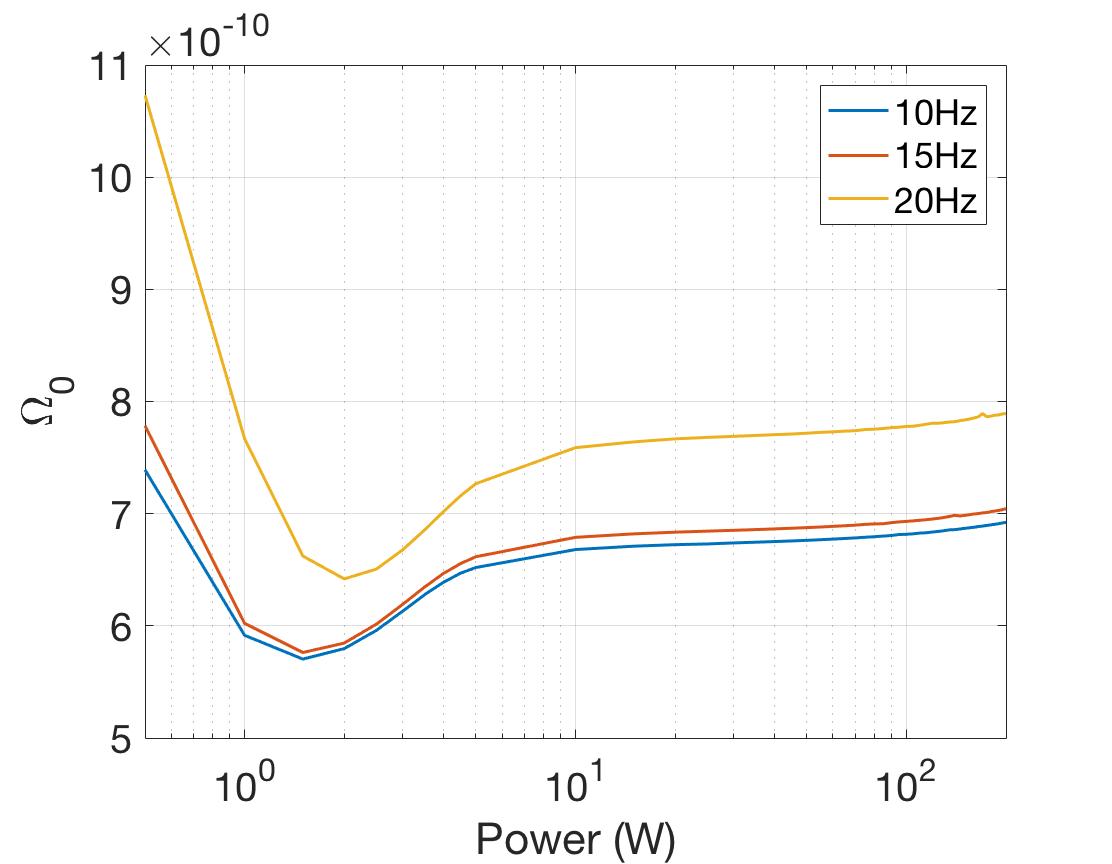}
	\caption{Plot of the SGWB sensitivity limit versus laser power and low-frequency cutoff. It can be seen that the best sensitivity for the SGWB is around $\sim 2$ W.}
	\label{fig:pow_omega}
\end{figure}

The relation between laser power, low-frequency cutoff and the optimal sensitivity $\Omega_{0}$ is presented in Fig.~\ref{fig:pow_omega}. These correspond to the optimal configurations found in Fig.~\ref{fig:pow_config}. As the power decreases, the sensitivity improves and is the best around 1.5 W. The optimal powers are 1.5 W for 10 Hz and 15 Hz cutoff, and 2 W for 20 Hz cutoff. For 10 Hz cutoff and 1.5 W laser power, the optimal $\Omega_{0}$ is $5.7\times 10^{-10}$, which is the global optimal sensitivity among all interferometer configurations considered. When the laser power is above 5 W, the sensitivity gets worse as the laser power increases. Assuming a 10 Hz cutoff, the optimal $\Omega_{0}$ is $6.5\times 10^{-10}$ at 5 W and it worsens (albeit not significantly) to $6.9\times 10^{-10}$ at 200 W. Also, reducing the low-frequency cutoff improves the sensitivity. For example, when we use a 125 W laser, a 10 Hz cutoff has a limit of $6.8\times 10^{-10}$ for $\Omega_{0}$ while a 20 Hz cutoff has $7.8\times 10^{-10}$. 

\begin{figure}[h]
	\includegraphics[width=0.45\textwidth]{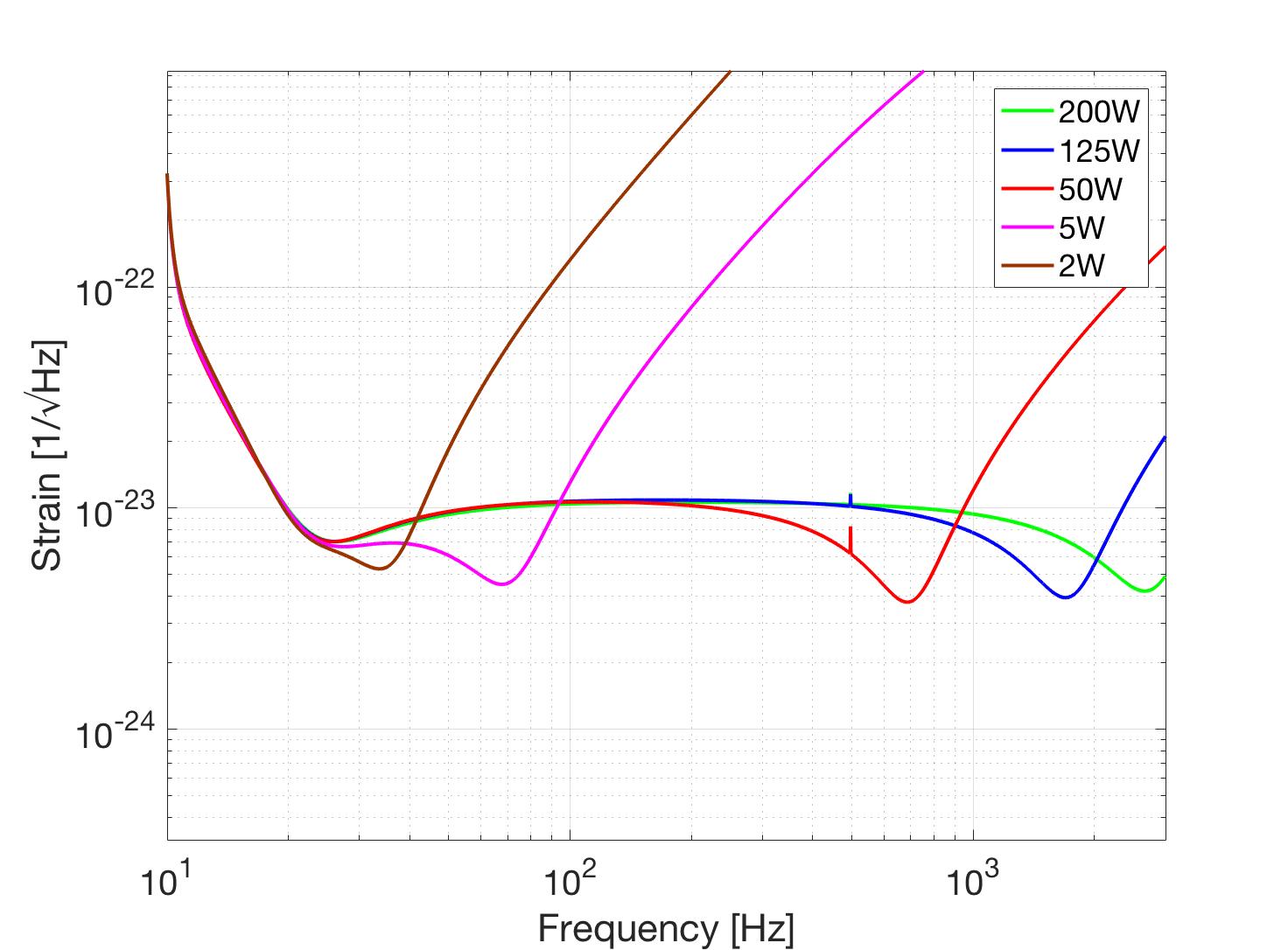}
	\caption{Dependence of the optimal SGWB noise spectra on laser power. The powers considered are 2 W, 5 W, 50 W, 125 W and 200 W, and the plotted spectra use optimal signal recycling configurations. With the decrease of laser power, the bottom part of the spectra becomes narrower, and the local minimum on the right shifts to lower frequencies. This low frequency improvement leads to a better sensitivity for the SGWB search.}
	\label{fig:pow_spec}
\end{figure}

In addition, another two observations can be made from the high power region of Fig.~\ref{fig:pow_omega}.
\begin{itemize}
\item The optimal sensitivity improves as power increases from zero until it reaches the optima at about 2 W. As the power keeps increasing from the optimal sensitivity, the sensitivity gets worse. However, we can see an interesting behavior in that $\Omega_{0}$ increases faster at the beginning and slows down later. The consequence of this is that, if we examine the increasing of $\Omega_0$ between 2 W and 200 W, we can see that most of the increase happens between 2 W and 10 W. The difference of $\Omega_{0}$ between 10 W and 200 W is less than the difference between 2 W and 10 W.
\item The relation between the low-frequency cutoff and the optimal $\Omega_{0}$ limit is non-linear. The gap between the 20 Hz line and the 15 Hz line in Fig.~\ref{fig:pow_omega} is much larger than the gap between the 15 Hz line and the 10 Hz line. This reflects the rapid change in the interferometer's sensitivity at low frequencies.
\end{itemize}
In order to look further into how the laser power affects sensitivity, we compare the noise spectra of some example laser powers in Fig.~\ref{fig:pow_spec}. The noise spectra for 2 W, 5 W, 50 W, 125 W and 200 W optimal configurations are plotted. One can see that the high frequency minimum shifts to the low frequency region as power decreases, lowering the noise at the low frequency region at the cost of high frequencies. Eventually at 5 W, both dips are below 100 Hz and this contributes to the low-frequency optimization. This explains the improvement of SGWB sensitivity when using low laser power. 

\begin{figure}[t]
    \centering
    \begin{subfigure}[t]{0.4\textwidth}
        \centering
        \includegraphics[width=\textwidth]{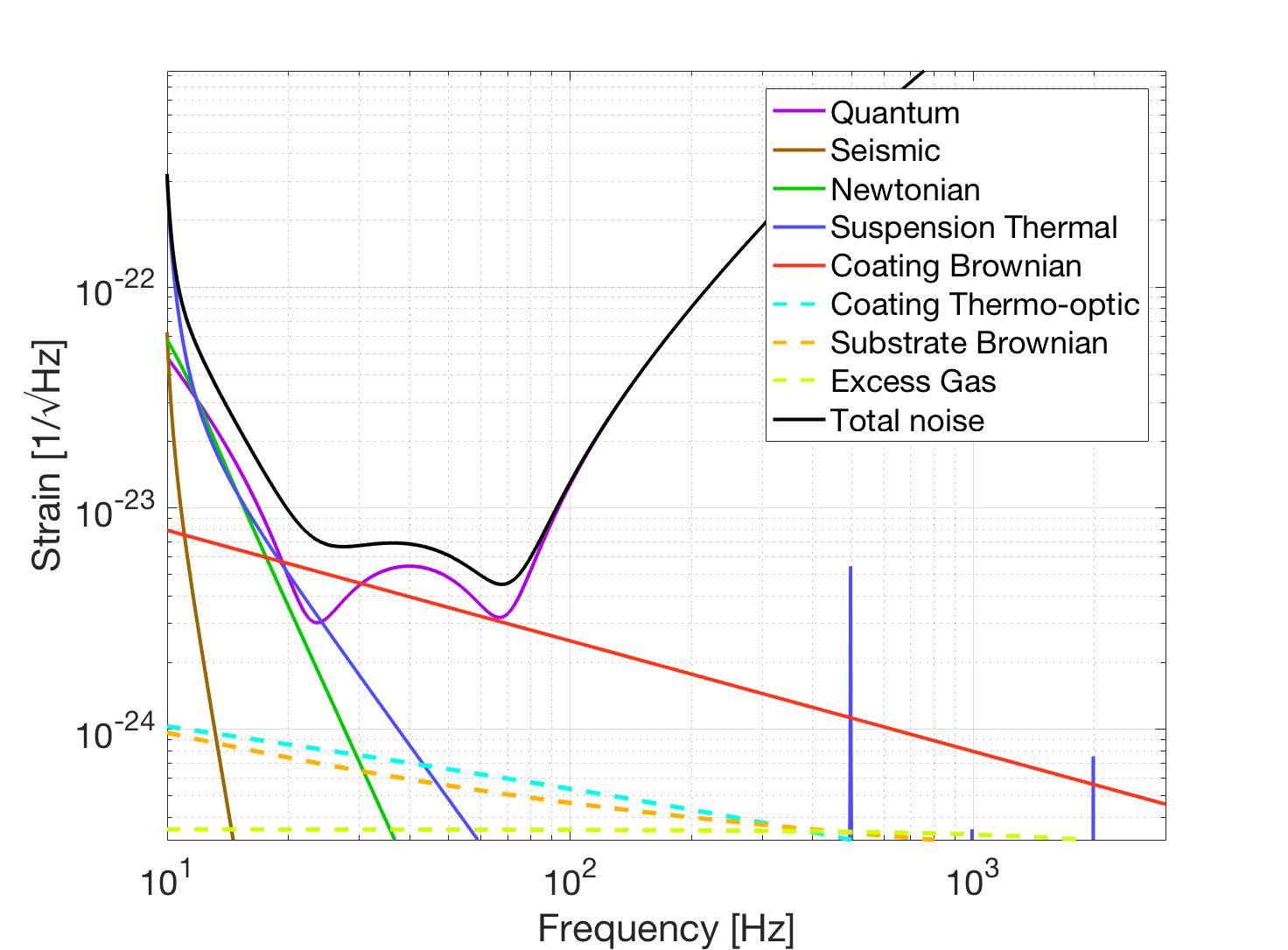}
    \end{subfigure}\\
    \begin{subfigure}[t]{0.4\textwidth}
        \centering
        \includegraphics[width=\textwidth]{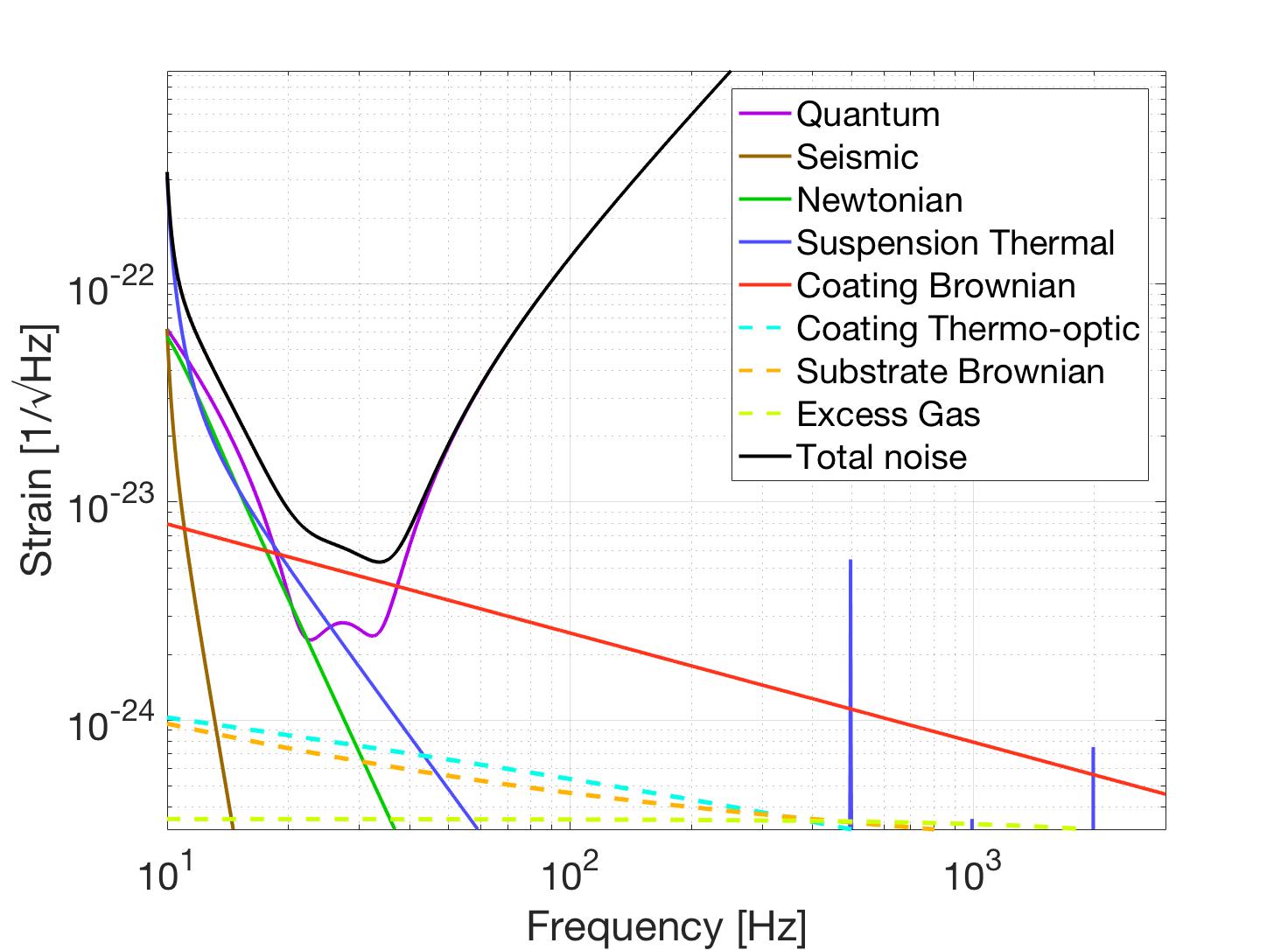}
    \end{subfigure}
    \caption{Optimal SRC configurations, and the contributing noise sources, for  5 W (top) and 2 W (bottom) of input laser power. These plots can be compared with Fig.~\ref{fig:opt_spectrum} for 125 W. For 5 W the quantum noise is dominant below 20 Hz, and then above 30 Hz. For 2 W of laser power the quantum noise has been reduced, and the Brownian noise for the mirror coatings dominates in the important low-frequency regime, in this case from 18 Hz to 38 Hz.}
    \label{fig:Low_pow_config}
\end{figure}

Contributing to the low frequency improvements is the reduction of the quantum noise (the sum of the radiation pressure noise and shot noise). Laser power is directly related to the radiation pressure noise~\cite{quantum_noise}, which dominates the quantum noise at low frequencies~\cite{quant_thesis}. Considering the significance of the low-frequency behavior of $h_n(f)$ we arrive at the point where lower laser power gives lower radiation pressure noise, which in turn leads to the low-frequency improvement shown, in Fig.~\ref{fig:pow_spec} and finally gives a better sensitivity limit as seen in Fig.~\ref{fig:pow_omega}. This is further displayed in Fig.~\ref{fig:Low_pow_config}. With the laser power at 5 W the quantum noise is dominant below 20 Hz, and then above 30 Hz. With 2 W of laser power the quantum noise has been reduced, and the Brownian noise for the mirror coatings dominates in the important low-frequency regime (18 Hz to 38 Hz in this example).

\section{Optimizing the search for a compact binary produced SGWB}
In Eq.~\ref{eq:pl_general}, we write a general model for the frequency dependence of $\Omega_{gw}(f)$ with two parameters $\alpha$ and the reference frequency $f_{ref}$. For all the work above, we assumed a flat SGWB model, $\alpha=0$, as is given in Eq.~\ref{eqn:bg_flat}. Here we explore a model that represents the background produced by compact binary mergers (binary black holes or binary neutron stars) over the history of the universe~\cite{power_law,Abbott:2017xzg}, where $\alpha = 2/3$ is assumed. 
Certainly there are proposed SGWBs where the frequency dependence is different~\cite{PhysRevLett.118.121101}, but we concentrate on the two models that are most likely to be detected by the advanced detector network in the coming years~\cite{Abbott:2017xzg}.
We will now optimize the signal recycling for a SGWB search with this $\alpha = 2/3$ frequency dependence and compare the results with the flat $\alpha=0$ model.

\subsection{The $\alpha=2/3$ model and sensitivity limit}
If we use $\alpha=2/3$ in Eq.~\ref{eq:pl_general}, we get
\begin{equation}
\label{eq:pl}
\Omega_{gw}(f) = \Omega_{2/3} (\frac{f}{f_{ref}})^{2/3}.
\end{equation}
We can represent the detection sensitivity limit with $\Omega_{2/3}$ and, according to Eq.\ref{eqn:lim_pl}, the sensitivity limit is
\begin{equation}\label{eqn:lim_pl2}
\Omega_{3/2}\geq\frac{25\pi c^2}{16\rho_c G}\sqrt{\frac{2}{T}}\Big[\int \frac{\gamma^2(\vec{x_1},\vec{x_2},f)}{h_n^4(f)f^{14/3}f_{ref}^{4/3}}df\Big]^{-1/2}.
\end{equation}
The use of $\alpha=3/2$ somewhat mitigates the fact that the low-frequency noise spectrum contributes more to the sensitivity limit than the high frequency spectrum with the $f^{-14/3}$ in the place of $f^{-6}$ in the integrand. The low-frequency spectrum is still much more important than that for higher frequencies.

\subsection{The optimization results}
\begin{figure}[t]
    \centering
    \begin{subfigure}[t]{0.4\textwidth}
        \centering
        \includegraphics[width=\textwidth]{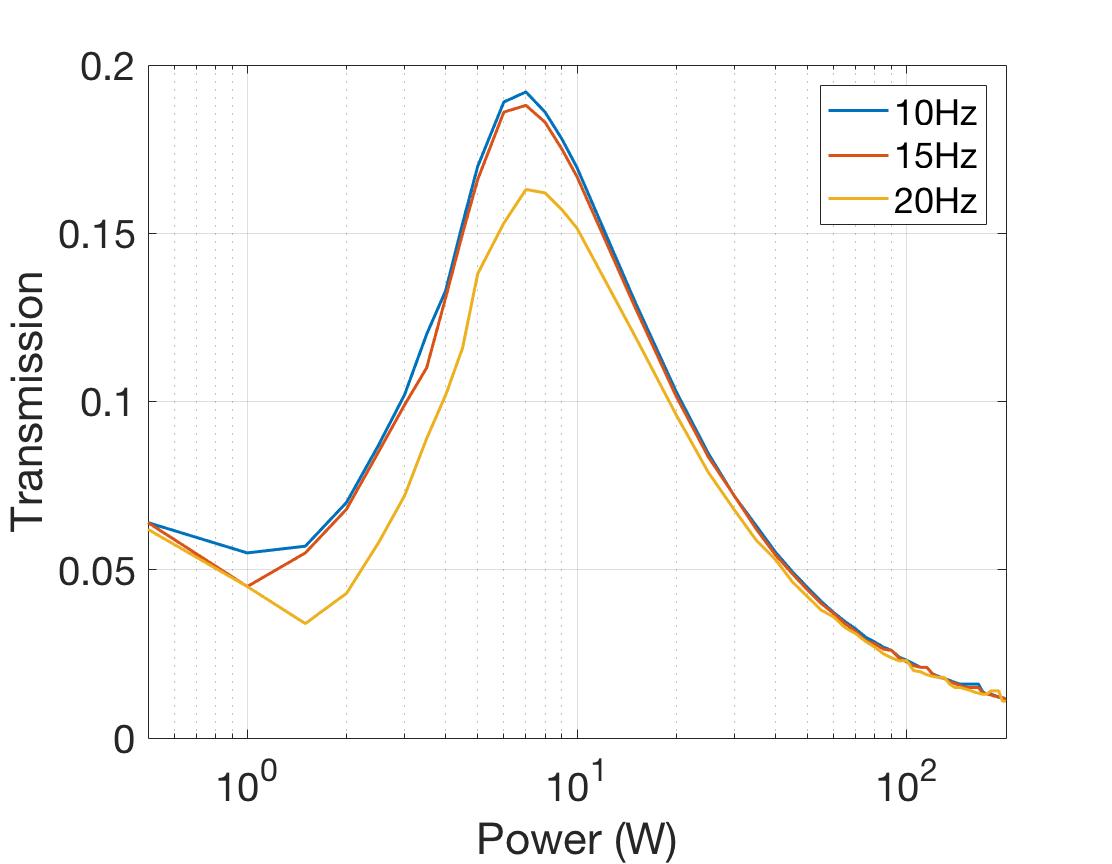}
    \end{subfigure}\\
    \begin{subfigure}[t]{0.4\textwidth}
        \centering
        \includegraphics[width=\textwidth]{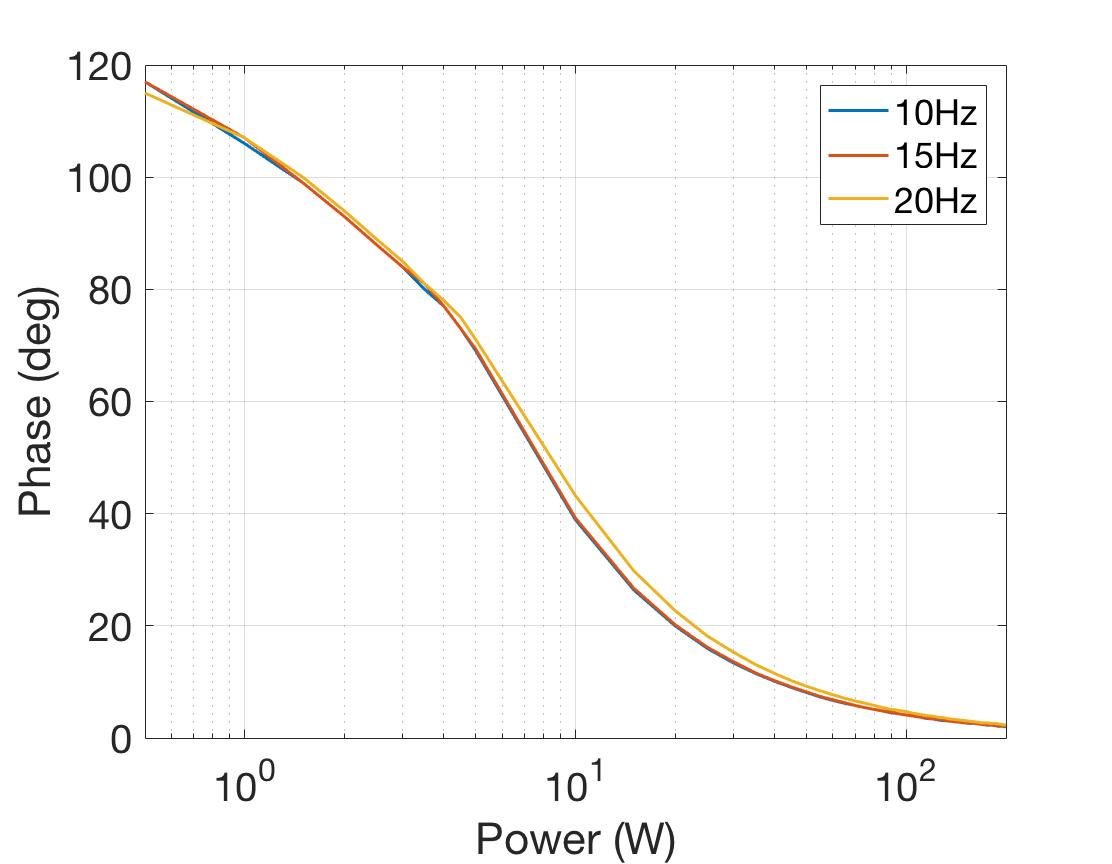}
    \end{subfigure}\\
    \begin{subfigure}[t]{0.4\textwidth}
        \centering
        \includegraphics[width=\textwidth]{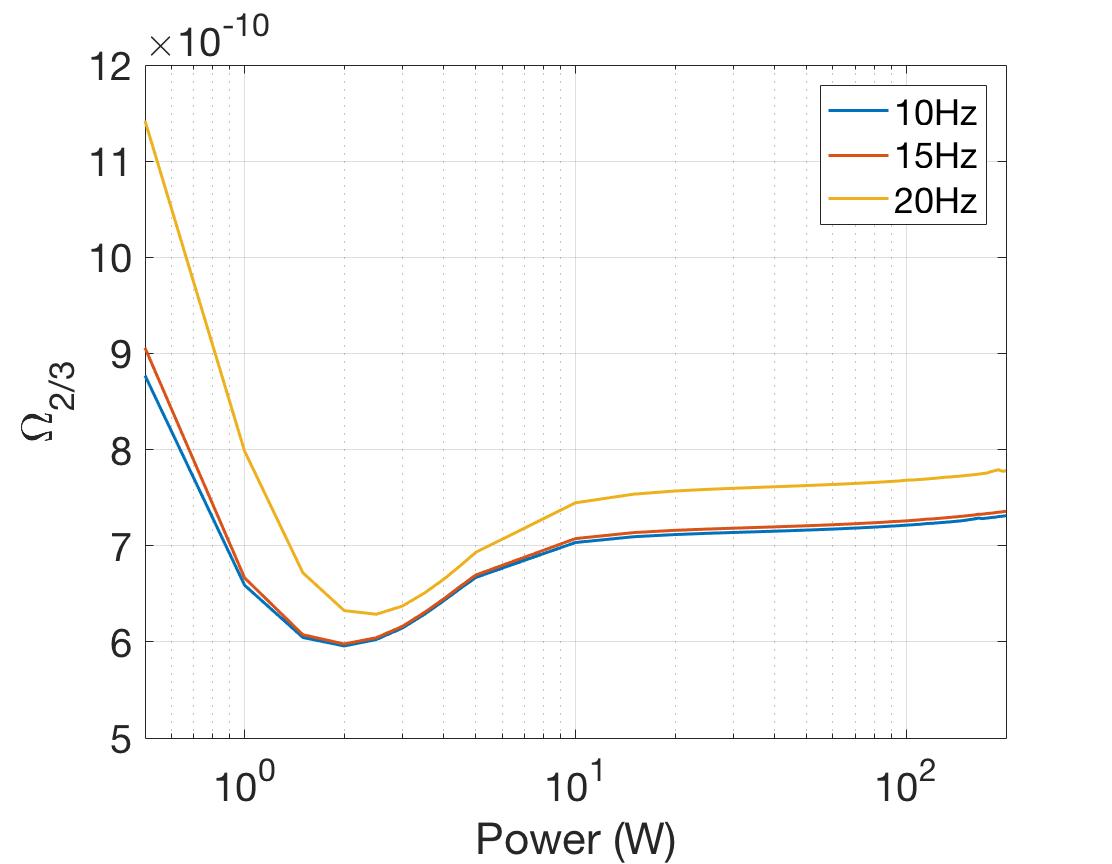}
    \end{subfigure}
    \caption{The optimization of the SGWB search assuming the an $\alpha = 2/3$ model given in Eq.~\ref{eq:pl}. On the top is the SRM transmission and in the middle is the signal recycling detuning phase. On the bottom is the limit on $\Omega_{2/3}$ as is given in Eq.~\ref{eqn:lim_pl} for different low-frequency cutoffs. The relations are all similar to the flat model shown in Fig.~\ref{fig:pow_config}.}
    \label{fig:pl_pow_config}
\end{figure}

Assuming $\alpha=2/3$, we find the optimal transmission and signal recycling detuning phase for every laser power usage. The results are shown in the top and middle plots of Fig.~\ref{fig:pl_pow_config}, which are very similar to the results in Fig.~\ref{fig:pow_config}. The optimal transmission initially increases with laser power, then decreases for higher powers. The optimal phase decreases with as the power increase. Therefore, the use  of the $\alpha=2/3$ power law model does not significantly affect the optimal configurations.

The optimal $\Omega_{2/3}$ as a function of laser power is shown in the bottom plot of Fig.~\ref{fig:pl_pow_config}. The relation is also similar to the flat model in Fig.~\ref{fig:pow_omega}, with $\Omega_{2/3}$ increasing with laser power or low-frequency cutoff above $\sim 2$ W and decreasing below. The similarity is not surprising considering that in both Eq.~\ref{eqn:bb_omega} and Eq.~\ref{eqn:lim_pl}, the low-frequency noise spectrum contributes more than the high frequency spectrum when $\alpha=2/3$.

\subsection{Comparisons of the $\alpha=0$ model and $\alpha=2/3$ models}
\begin{figure*}
    \centering
    \begin{subfigure}[t]{0.33\textwidth}
        \centering
        \includegraphics[width=\textwidth]{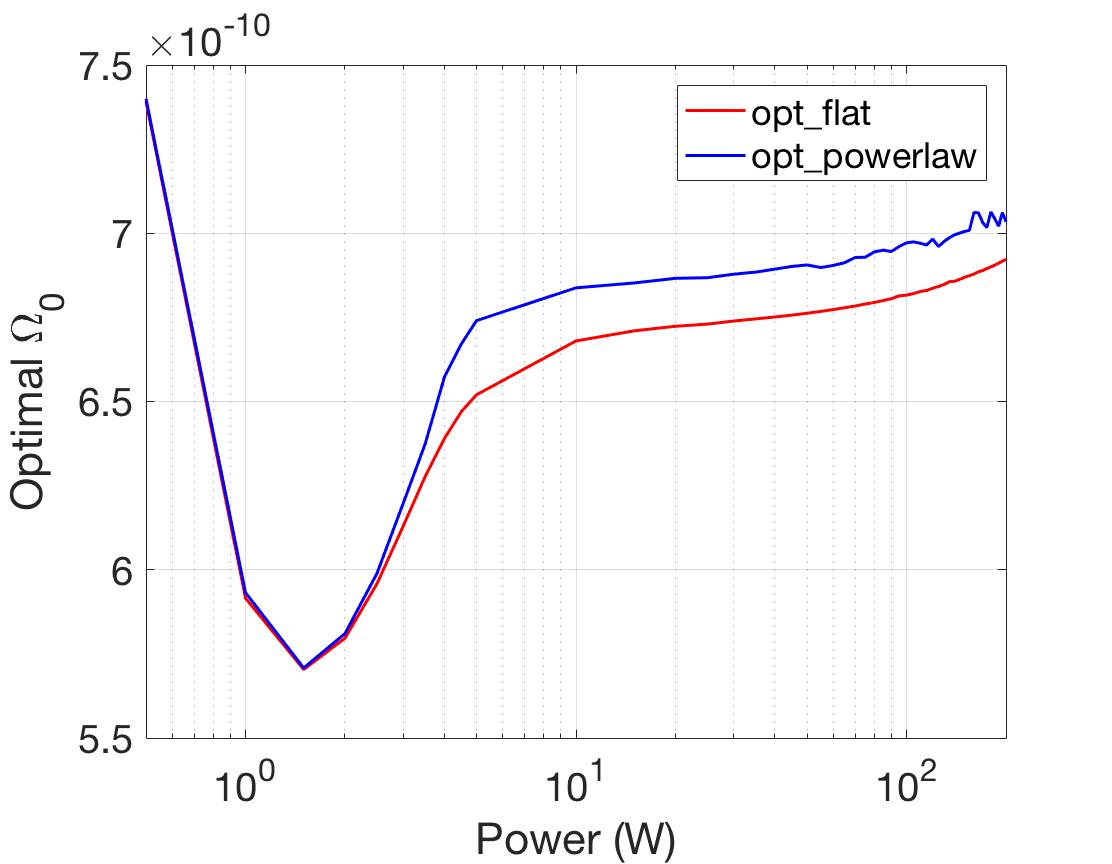}
    \end{subfigure}%
    \begin{subfigure}[t]{0.33\textwidth}
        \centering
        \includegraphics[width=\textwidth]{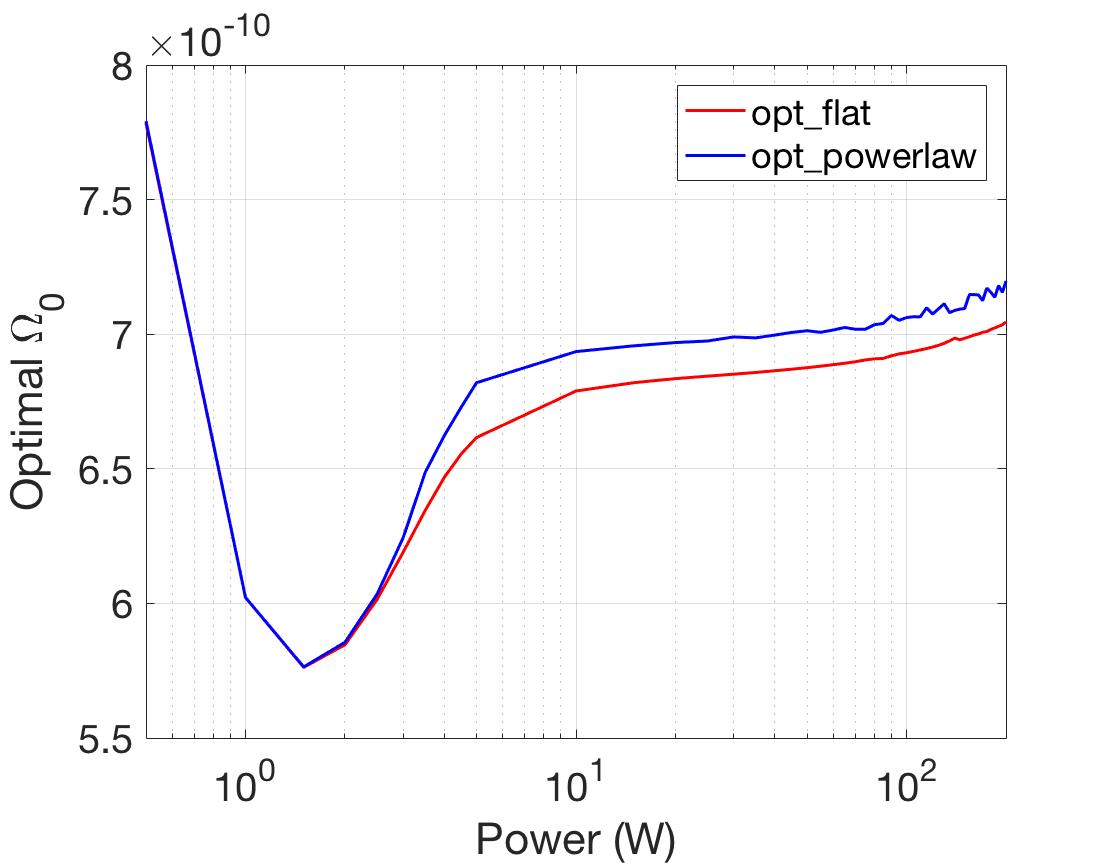}
    \end{subfigure}
    \begin{subfigure}[t]{0.33\textwidth}
        \centering
        \includegraphics[width=\textwidth]{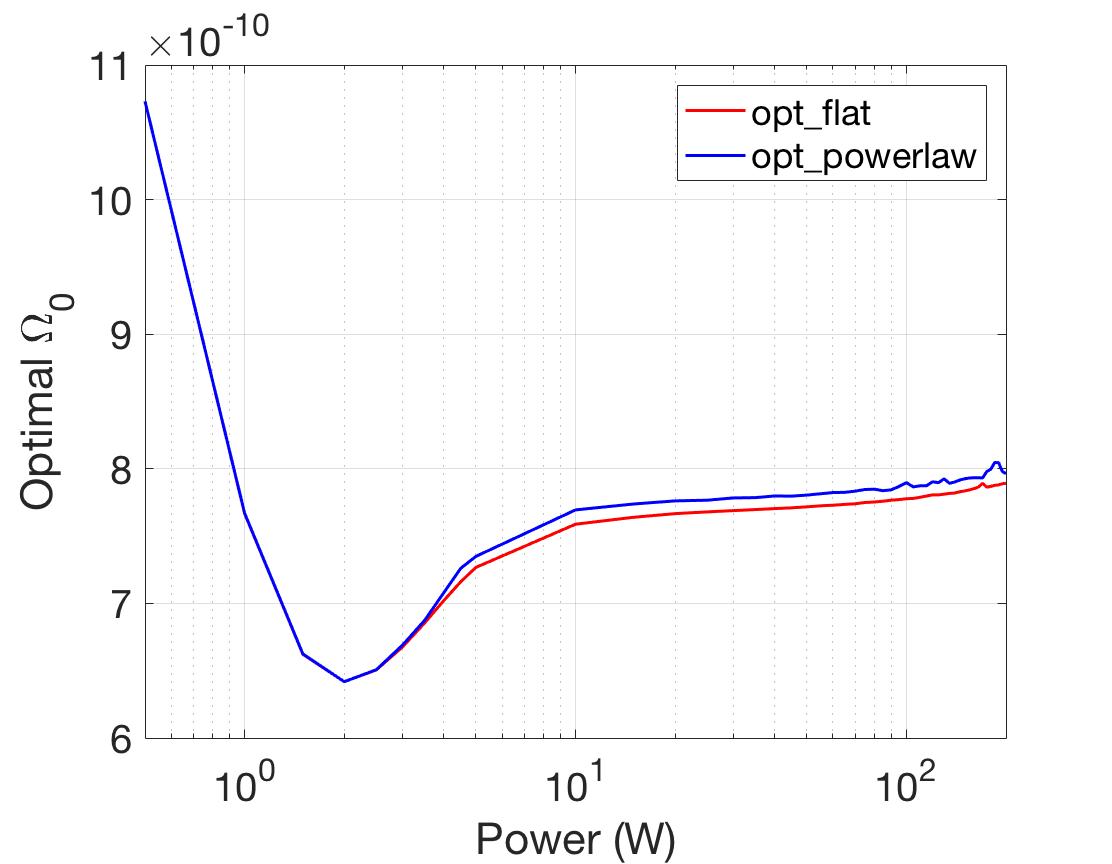}
    \end{subfigure}
    \begin{subfigure}[t]{0.33\textwidth}
        \centering
        \includegraphics[width=\textwidth]{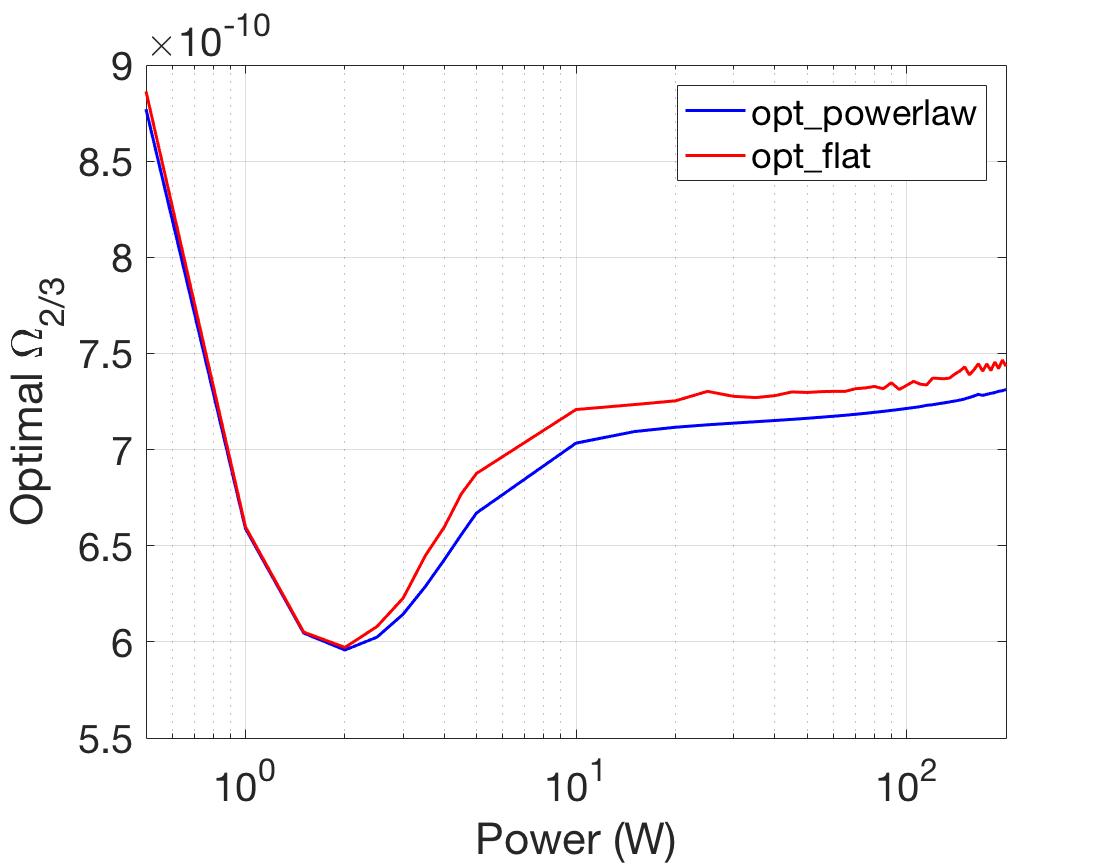}
    \end{subfigure}%
    \begin{subfigure}[t]{0.33\textwidth}
        \centering
        \includegraphics[width=\textwidth]{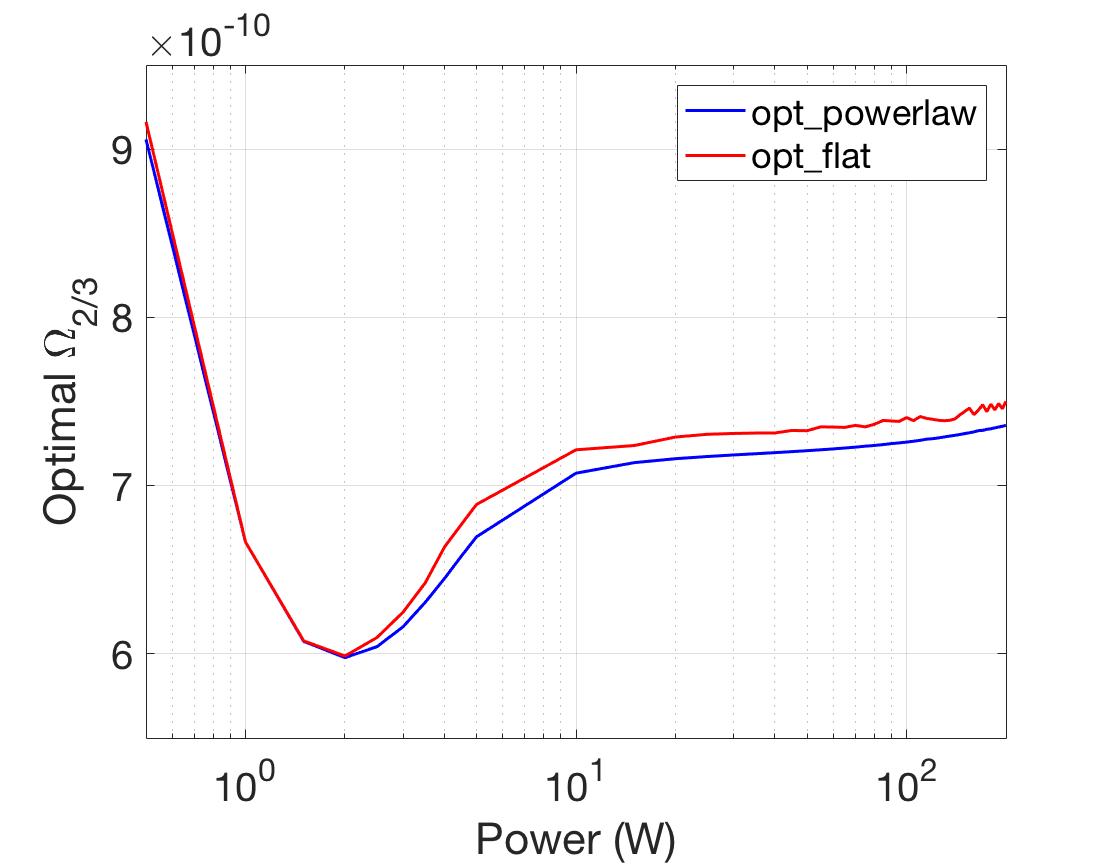}
    \end{subfigure}
    \begin{subfigure}[t]{0.33\textwidth}
        \centering
        \includegraphics[width=\textwidth]{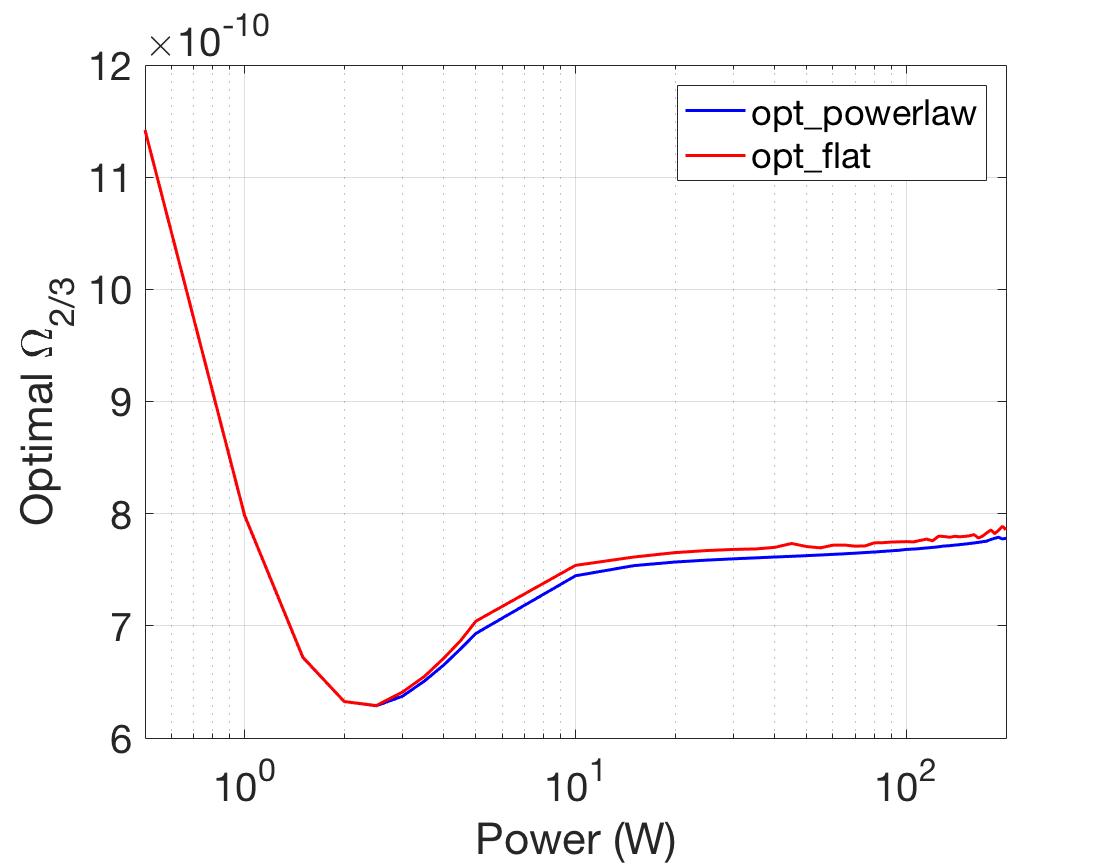}
    \end{subfigure}
    \caption{Comparisons of the sensitivities optimized for the $\alpha = 2/3$ power law and the flat $\alpha = 0$ SGWB models. On the top, left, middle, right are 10 Hz, 15 Hz, 20 Hz comparisons assuming the flat model. Bottom is the same assuming the power law model. Based on the comparisons, it is found that an optimal configuration for one model has a sensitivity close to the optimal sensitivity in another model.}
    \label{fig:model_comp}
\end{figure*}
The work above optimizes signal recycling for the SGWB search using both the flat $\alpha=0$ model and the power law $\alpha=2/3$ model (see Eq.~\ref{eq:pl_general}). In this section, we investigate how the flat optimal configurations perform when the power law model is assumed and, similarly, how the power law optimal configurations perform when the flat model is assumed.

For each set of the optimal configurations of the flat model, we compute its $\Omega_{2/3}$ limit given as Eq.~\ref{eqn:lim_pl}. Also, for each set of optimal configurations for the power law model, we compute its $\Omega_0$ limit given as Eq.~\ref{eqn:bb_omega}. We have comparisons for each frequency cutoff of 10 Hz, 15 Hz and 20 Hz. The comparison plots are shown in Fig.~\ref{fig:model_comp}.

The plots on the first row of Fig.~\ref{fig:model_comp} show the $\Omega_0$ limit computed using the $\alpha=0$ model in Eq.\ref{eqn:bb_omega} as a function of laser power, and the plots on the second row shows $\Omega_{2/3}$ limit computed using the $\alpha=2/3$ model in Eq.\ref{eqn:lim_pl2}. The plots in the left, middle and right columns assume 10 Hz, 15 Hz and 20 Hz low-frequency cutoff respectively. In all the plots, we compare two sets of configurations: configurations optimized for the $\alpha=0$ model (red line) and the configurations optimized for the $\alpha=2/3$ model (blue line). It can be seen in all the six plots in Fig.~\ref{fig:model_comp} that the gaps between the lines are approximately $10^{-11}$. Therefore, the sensitivity limits achieved by either optimal configurations are close to each other, so we can use the configurations optimized for either model to detect the SGWB.

\section{Conclusion}
The SRM transmission and signal recycling detuning phase are optimized for the SGWB  search with different laser powers, low-frequency cutoffs and models for the frequency dependence of $\Omega_{gw}(f)$. For a 125 W laser and a low-frequency cutoff at 10 Hz, the optimal transmission is found to be $1.5 \pm 0.01\%$ and the optimal phase $2.7 \pm  0.1^\circ$, giving a limit on $\Omega_{0}$ of $6.8 \times 10^{-10}$ for a year of observation time. The sensitivity for the SGWB search and BNS search for four configurations are compared. It is confirmed that a signal recycled interferometer system is more sensitive for SGWB and BNS searches. Also, the BNS optimized configuration is found to have relatively good sensitivity for both the SGWB search and the BNS search. However, we subsequently found that it is difficult to achieve better sensitivities for both the SGWB and BNS searches. 

There might be some ways to address the compatibility issue of the BNS search and the SGWB searches, which could lead to potential future studies. First, we might consider replacing the SRM to produce different transmission and detuning phase for different tasks. However, that would be simple in our simulation studies but could lead to more on-site engineering challenges. On the other hand, if we cannot change the configurations very easily, the best we can do is to find a trade-off between the BNS search and the SGWN search. For example, we can define a metric involving both $\Omega_{gw}$ and the BNS range. Then, we scan the SRM transmission and detuning phase to optimize the metric, in order to achieve a balance between the two sensitivities.

The optimization is then generalized to interferometers with different laser powers and frequency cutoffs. It turns out that using lower laser powers above 2 W results in higher optimal transmission, higher optimal signal recycling detuning phase and better sensitivity for a SGWB search. Besides, the cutoff frequency does not have a significant effect on the optimal configurations but a lower cutoff gives better sensitivity. The best sensitivity for a SGWB was found with a laser power of 1.5 W and a 10 Hz low-frequency cutoff, giving a limit on $\Omega_{0}$ of $5.7 \times 10^{-10}$ for a year of observation time.

Finally, we consider the SGWB produced by all the compact binary mergers in the history of the universe. This gives us a power law model given in Eq.~\ref{eq:pl}. The  $\alpha = 2/3$ power law model does not significantly change the optimization. Actually, configurations optimized for the flat $\alpha = 0$ model have sensitivities that are close to the optimal sensitivities for the $\alpha = 2/3$ power law model. This would mean that an optimized configuration, either for the flat model or the power law model, would have good sensitivities for both.

In the future, one might envision implementing even more complicated signal recycling systems. The addition of an internal signal recycling mirror between the SRM and the beamsplitter and an optomechanical filter module were recently proposed to achieve a broadband resonance~\cite{miao2017towards}. Besides, different models for a SGWB can also be considered. Gravitational-wave emission from the BNS post-merger remnant is expected to produce a SGWB above 1 kHz~\cite{miao2017towards}. It will be of scientific interest to look further into the optimization of more complicated signal recycling systems for different models of the SGWB above 1 kHz.
\\
\section*{acknowledgements}
This work is supported by National Science Foundation (NSF) grant PHY-1505373 to Carleton College. We would also like to thank Peter Fritschel for kindly providing feedback and suggestions during multiple stages of this work. Vuk Mandic and Xingjiang Zhu also provided useful comments.
\bibliographystyle{unsrt}
\bibliography{ref}
\end{document}